\documentclass[journal,letter]{IEEEtran}						

\usepackage{cite}
\usepackage[cmex10]{amsmath} 

\usepackage{amssymb}
\usepackage{amsthm}	
\usepackage{color}
\usepackage{tikz}	
\usetikzlibrary{backgrounds,fit,arrows,decorations.pathreplacing,positioning}
\usepackage{algorithm}
\usepackage{extarrows}
\usepackage{dblfloatfix}	
\usepackage{subfigure} 		

\usepackage[normalem]{ulem}	

\usepackage{xargs}
\usepackage[colorinlistoftodos,prependcaption]{todonotes}

\newcommandx{\rednote}[2][1=]{\todo[linecolor=red,backgroundcolor=red!25,bordercolor=red,#1]{#2}}
\newcommandx{\bluenote}[2][1=]{\todo[linecolor=blue,backgroundcolor=blue!25,bordercolor=blue,#1]{#2}}
\newcommandx{\yellownote}[2][1=]{\todo[linecolor=yellow,backgroundcolor=yellow!25,bordercolor=yellow,#1]{#2}}
\newcommandx{\greennote}[2][1=]{\todo[inline,linecolor=olive,backgroundcolor=green!25,bordercolor=olive,#1]{#2}}

\newtheorem{definitionenv}{Definition}
\newtheorem{lemmaenv}[definitionenv]{Lemma}
\newtheorem{theoremenv}[definitionenv]{Theorem}
\newtheorem{corollaryenv}[definitionenv]{Corollary}
\newtheorem{propositionenv}[definitionenv]{Proposition}
\newtheorem{conjectureenv}[definitionenv]{Conjecture}

\newtheorem{app-lemmaenv}[section]{Lemma}
\newtheorem{remarkenv}[definitionenv]{Remark}

\newenvironment{definition}{\begin{definitionenv}\rm}{\end{definitionenv}}
\newenvironment{lemma}{\begin{lemmaenv}\rm}{\end{lemmaenv}}
\newenvironment{theorem}{\begin{theoremenv}\rm}{\end{theoremenv}}
\newenvironment{corollary}{\begin{corollaryenv}\rm}{\end{corollaryenv}}
\newenvironment{proposition}{\begin{propositionenv}\rm}{\end{propositionenv}}
\newenvironment{conjecture}{\begin{conjectureenv}\rm}{\end{conjectureenv}}
\newenvironment{app-lemma}{\begin{app-lemmaenv}\rm}{\end{app-lemmaenv}}
\newenvironment{remark}{\begin{remarkenv}\rm}{\end{remarkenv}}

\newcommand{\bd}{\begin{definition}}
\newcommand{\ed}{\end{definition}}
\newcommand{\edp}{\hspace*{\fill} $\Box$ \end{definition}}
\newcommand{\bl}{\begin{lemma}}
\newcommand{\el}{\end{lemma}}
\newcommand{\elp}{\hspace*{\fill} $\Box$ \end{lemma}}
\newcommand{\bt}{\begin{theorem}}
\newcommand{\et}{\end{theorem}}
\newcommand{\etp}{\hspace*{\fill} $\Box$ \end{theorem}}
\newcommand{\bc}{\begin{corollary}}
\newcommand{\ec}{\end{corollary}}
\newcommand{\ecp}{\hspace*{\fill} $\Box$ \end{corollary}}
\newcommand{\bcj}{\begin{conjecture}}
\newcommand{\ecj}{\end{conjecture}}
\newcommand{\be}{\begin{example}}
\newcommand{\ee}{\end{example}}
\newcommand{\eep}{\hspace*{\fill} $\Box$ \end{example}}
\newcommand{\bp}{\begin{proposition}}
\newcommand{\ep}{\end{proposition}}
\newcommand{\epp}{\hspace*{\fill} $\Box$ \end{proposition}}
\newcommand{\br}{\begin{remark}}
\newcommand{\er}{\end{remark}}
\newcommand{\erp}{\hspace*{\fill} $\Box$ \end{remark}}

\DeclareMathOperator{\sgn}{sign}
\DeclareMathOperator*{\argmax}{arg\,max}

\newcommand{\eq}[1]{\eqref{#1}}       

\newcommand{\mbf}[1]{\pmb{#1}}

\newcommand{\beq}[1]{\begin{equation*}{ #1 }\end{equation*}}
\newcommand{\beql}[2]{\begin{equation}\label{#1}{ #2 }\end{equation}}
\newcommand{\beqs}[1]{\begin{align*} #1 \end{align*}}

\newcommand{\bcase}[1]{\begin{cases} #1 \end{cases}}

\newcommand{\bsmtx}[1]{\left[ \begin{smallmatrix} #1 \end{smallmatrix} \right]}

\newcommand{\sC}{{\cal C}}

\newcommand{\sM}{{\cal M}}
\newcommand{\sN}{{\cal N}}

\newcommand{\af}{\alpha}
\newcommand{\dlt}{\delta}

\newcommand{\La}{\Lambda}

\newcommand{\eps}{\epsilon}

\newcommand{\ps}[1]{\left( {#1} \right)}

\begin{document}
%
\title{Refined Belief Propagation Decoding of Sparse-Graph Quantum Codes}

%
\author{Kao-Yueh~Kuo 
    and Ching-Yi~Lai
\thanks{K.-Y.~Kuo and C.-Y.~Lai are with the Institute of Communications Engineering, National Chiao Tung University, Hsinchu 30010, Taiwan. 
	(e-mail: kykuo@nctu.edu.tw and cylai@nctu.edu.tw)}
}



\maketitle

\begin{abstract}
 
Quantum stabilizer codes constructed from sparse matrices have good performance and can be efficiently decoded by belief propagation (BP).
A conventional BP decoding algorithm treats binary stabilizer codes as additive codes over GF(4). 
This algorithm has a relatively complex process of handling  check-node messages, which incurs
higher decoding complexity. 
Moreover, BP decoding of a stabilizer code usually suffers a performance loss due to the many short cycles in the underlying Tanner graph.
In this paper, we propose a refined BP decoding algorithm for quantum codes with complexity roughly the same as binary BP. 
For a given error syndrome, this algorithm decodes to the same output as the conventional quaternary BP but the passed  {node-to-node} messages are single-valued, unlike the quaternary BP, where multivalued  {node-to-node} messages are required.
Furthermore, the techniques of message strength normalization can naturally be applied to these single-valued messages to improve the performance.
Another observation is that the message-update schedule affects the performance of BP decoding  against short cycles. 
We show that running BP with message strength normalization according to a serial schedule (or other schedules) may significantly improve the
decoding performance and error floor in computer simulation.
\end{abstract}

\begin{IEEEkeywords}
Quantum stabilizer codes, LDPC codes, sparse matrices, belief propagation, sum-product algorithm, decoding complexity and performance, message-update schedule, message normalization.
\end{IEEEkeywords}

%
\IEEEpeerreviewmaketitle

\section{Introduction} \label{sec:Intro}
%
In classical coding theory, low-density parity-check (LDPC) codes and the sum-product (decoding) algorithm, proposed by Gallager, are shown to have near Shannon-capacity performance for the binary symmetric channel (BSC) and the additive white Gaussian noise  {(AWGN)} channel \cite{Gal62,Gal63,MN96,Mac99}. 
The sum-product algorithm is understood as a \textit{message-passing} algorithm running on the Tanner graph~\cite{Tan81} 
corresponding to a parity-check matrix of the underlying linear code.
This is also known as a realization of Pearl's belief propagation (BP) algorithm \cite{Pea88,MMC98,YFW03}.
It is efficient~\cite{Wib96,AM00,KFL01}
and its complexity is roughly proportional to the number of edges in the Tanner graph \cite{Wib96,DM98}. 
The messages or beliefs are real values that can be used to infer the conditional probabilities of the variables from the observed signals (e.g., error syndromes).  
The BP algorithm will have \emph{variable-to-check} and \emph{check-to-variable} messages passed around the Tanner graph according to a predefined schedule~\cite{Wib96}.
Commonly used schedules include the parallel (flooding) schedule and the serial (sequential/shuffled) schedule~\cite{ZF05,SLG07,GH08}.
In practice, an important topic is to discuss message approximation and quantization with necessary compensations such as normalization and offset~\cite{CF02b,CDE+05,YHB04}.
BP can be interpreted as a gradient descent algorithm \cite{LBB98}, and the message normalization/offset is a strategy 
to adjust the update step-size so that the effects of short cycles in the Tanner graph can be mitigated~\cite{JN06,EM06}.

The idea of error correction has been applied to protect quantum information against noise. Quantum error-correcting codes, especially the class of quantum stabilizer codes, bear similarities to classical codes~\cite{Shor95,GotPhD}.
Calderbank, Shor, and Steane (CSS) showed that good stabilizer codes can be constructed from classical dual-containing codes \cite{CS96,Steane96}. 
MacKay, Mitchison, and McFadden proposed  various methods to build quantum LDPC codes  from self-orthogonal sparse matrices using the CSS construction~\cite{MMM04}, and they found several good quantum codes.
In particular, bicycle codes are of particular interest because of their good performance and low decoding complexity with BP~\cite{MMM04,PC08,Wan+12,Bab+15,ROJ19,PK19}.
There are a variety of sparse stabilizer codes constructed \cite{Kit03,COT05,HI07,Djo08,Aly08,KHIS11,TZ14,TL09,CDZ13,KP13}.

We will focus on quantum information in qubits in this paper. 
The error discretization theorem \cite{NC00} allows us to focus on a set of discrete error operators and we consider errors that are tensor product of Pauli matrices $I,X,Y,$ and~$Z$.
In this case, decoding a stabilizer code is equivalent to decoding an {additive} code over GF(4) 
with a binary error syndrome vector \cite{CRSS98,MMM04,PC08}.

In non-binary BP  over GF($q$), a message is a vector of $q$ real numbers that represent a distribution of the elements in GF($q$). 
The complexity for generating a variable-to-check message is simply $O(q)$ per edge, 
but the complexity  for generating a check-to-variable message is $O(q^2)$ per edge\,\footnote{ 
The complexity can be reduced to $O(q\log q)$ if the fast Fourier transform (FFT) method is appropriately used. Herein, we do not consider this method since there is additional cost running FFT and inverse FFT.}~\cite{DM98,DF07}.
In contrast, for binary BP a scalar message suffices to represent a binary distribution  $(p^{(0)},p^{(1)})$.
For example, 
 the \emph{likelihood difference} (LD) $\dlt = p^{(0)} - p^{(1)}$ and the \emph{log-likelihood ratio} (LLR) $\La = \ln(p^{(0)}/p^{(1)})$ are usually used
 and the corresponding BP update rules are called $\delta$-rule and $\Lambda$-rule, respectively
\cite{Gal63,Mac99}, \cite[Sec. V.E]{KFL01}.\,\footnote{
	These rules are mathematically equivalent.
	While the $\Lambda$-rule is often chosen for the AWGN channel~\cite{Gal63},
	the $\dlt$-rule is  suitable for the BSC, 
		as well as for our purpose of decoding quantum codes as shown in Sec.~\ref{sec:QuanBP}. 
	We will follow (47)--(53) in \cite{Mac99} to implement the $\dlt$-rule. 
	}
As a result, the BP decoding algorithm for a classical binary code runs 16 times faster than the quaternary BP algorithm for a quantum stabilizer code of comparable size.
To reduce the quantum decoding complexity, a common strategy is to treat a binary stabilizer code as a binary classical code with doubled length~\cite{MMM04,Bab+15}, 
followed by additional processes to handle the $X$/$Z$ correlations \cite{DT14,ROJ19}.

On the other hand,  a stabilizer code inevitably has many four-cycles in its Tanner graph, which degrade the performance of BP.
To overcome this problem, additional processes are proposed, such as heuristic flipping from nonzero syndrome bits \cite{PC08}, 
	(modified) enhanced-feedback \cite{Wan+12,Bab+15}, BP-based neural network \cite{LP19}, 
	augmented decoder (adding redundant rows to the parity-check matrices) \cite{ROJ19} , and ordered statistics decoding (OSD) \cite{PK19}.

In this paper, we simplify and improve the conventional quaternary BP for decoding a binary stabilizer code. Instead of passing  multivalued messages (corresponding to $q=4$ components) on the edges of the Tanner graph  {(see Fig.~\ref{fig:S2x3} for an example)}, we show that it is sufficient to pass \emph{single-valued} messages. 
An important observation is that the error syndrome of a binary stabilizer code, which can be considered as a quaternary code, is binary.
More precisely, a syndrome bit represents the commutation relation of the error and a stabilizer generator.
Consequently, a message from one node to another should reflect whether an error component is more likely to commute or anticommute with the stabilizer component.
Inspired by  MacKay's $\dlt$-rule for message passing in binary linear codes \cite{Mac99}, 
we derive a $\dlt$-rule based BP decoding algorithm for stabilizer codes.  
That is, the likelihood difference of the operators that commute or anticommute with the underlying stabilizer component is passed as a message.
This greatly improves the efficiency of BP.
Moreover, each of the previously-mentioned processes for BP improvement can be incorporated  in our algorithm.

To improve the performance while having low complexity for BP decoding of quantum codes, we have found two particularly useful methods. 
First, running the BP decoding with a serial schedule \cite{ZF05,SLG07} can improve the convergence behavior when the underlying Tanner graph has many short cycles. 
For illustration, we decode the $[[5,1]]$ code \cite{LMPZ96} and a $[[129,28]]$ hypergraph-product code with both serial and parallel schedules. 
In the case of $[[5,1]]$ code, the BP decoding converges quickly using the serial schedule, while it diverges using the parallel schedule. The serial schedule also outperforms the parallel schedule in the case of the $[[129,28]]$ hypergraph-product code.
Second, adjusting message magnitudes can improve the error floor performance~\cite{CF02b,CDE+05,YHB04}. 
The techniques of message normalization and offset are simple and efficient for improving the decoding performance. In addition, both  check-to-variable and variable-to-check messages can be separately adjusted. 
However, these techniques have not been considered in the quantum scenario probably because they are designed for binary BP decoding.
In our $\delta$-rule based BP, these techniques can be directly applied to the single-valued messages.
Moreover, all the mentioned techniques  
can be simultaneously applied in our BP algorithm. We have tested several quantum codes 
and the simulation results show that the decoding performance and error floor are improved significantly.

This paper is organized as follows. 
In Sec.~\ref{sec:ClsBP} we define the notation and review the binary BP decoding with two update schedules.
In Sec.~\ref{sec:QuanBP} we show how to efficiently compute the quaternary BP decoding for quantum stabilizer codes 
by single-valued message-passing.  
In Sec.~\ref{sec:NorQBP} we review the message normalization and offset algorithms, and port them to our decoding procedure. 
Related simulations are   provided.
Finally, we conclude in Sec.~\ref{sec:Conclu}. 

\section{Classical Binary Belief Propagation Decoding} \label{sec:ClsBP}

Consider a classical binary $[N,K]$ linear code $\sC$, defined by an $M \times N$ parity-check matrix $H\in\{0,1\}^{M\times N}$ (not necessarily of full rank) with $M\geq N-K$.
Suppose that a message is encoded by $\sC$ and sent through a noisy channel.
The noisy channel will introduce an $N$-bit random error vector $E=(E_1,E_2,\dots, E_N)$ corrupting the transmitted codeword. 
(A string $(X_1,X_2,\dots, X_N)$ is understood as a column vector in this paper.) 
Given an observed error syndrome vector $z\in\{0,1\}^M$,
the decoding problem of our concern is to find the most likely error vector $e^*\in\{0,1\}^N$ such that $He^*=z \mod 2$. (From now on the modulo operation will be omitted without confusion.)
More precisely, the maximum likelihood decoding is to find 
$$e^* = \argmax_{e\in\{0,1\}^N,\, He = z} P(E=e|z),$$
where  $P(E=e|z)$ is the probability of channel error $e$ conditioned on the observed syndrome $z$.
The above decoding problem can be depicted as a \textit{Tanner graph} 
and an approximate solution can be obtained by \textit{belief propagation} (BP) on the  Tanner graph. 

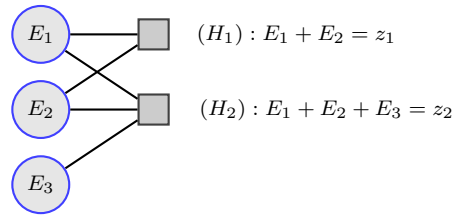
\begin{figure}[htbp] \centering ~~~~~~~~ \begin{tikzpicture}[node distance=1.3cm,>=stealth',bend angle=45,auto]

\tikzstyle{chk}=[rectangle,thick,draw=black!75,fill=black!20,minimum size=4mm]
\tikzstyle{var}=[circle,thick,draw=blue!75,fill=gray!20,minimum size=4mm,font=\footnotesize]
\tikzstyle{VAR}=[circle,thick,draw=blue!75,fill=blue!20,minimum size=5mm,font=\footnotesize]
\tikzstyle{fac}=[anchor=west,font=\footnotesize]

\node[var] (x3) at (0,0) {$E_3$};
\node[var] (x2) at (0,1) {$E_2$};
\node[var] (x1) at (0,2) {$E_1$};
\node[chk] (c1) at (1.5,2) {};
\node[chk] (c2) at (1.5,1) {};

\draw[thick] (x1) -- (c1) -- (x2);
\draw[thick] (x1) -- (c2) -- (x3);
\draw[thick] (x2) -- (c2);


\node[fac] [right of=c1,xshift=6mm] {$(H_1): E_1+E_2=z_1$};
\node[fac] [right of=c2,xshift=10mm] {$(H_2): E_1+E_2+E_3=z_2$};


\end{tikzpicture}
	\caption{The Tanner graph of   $H=\left[{1\atop 1}{1\atop 1}{0\atop 1}\right]$. The two squares are check nodes and the three circles are variable nodes.} \label{fig:H2x3}
\end{figure}

The Tanner graph corresponding to $H$ is a bipartite graph consisting of $N$ variable nodes and $M$ check nodes, 
and it has an edge connecting check node $m$ and variable node $n$ if the entry $H_{mn}=1$.
For our purpose, variable node $n$ corresponds to the random error bit $E_n$ and check node $m$ corresponds to a parity check $H_m$.
An example of $H=\begin{bmatrix}1 &1 &0\\ 1 &1 &1\end{bmatrix}$ is shown in Fig.~\ref{fig:H2x3}. 

To infer $e^*$, a BP algorithm computes an approximated marginal distribution $\hat P(E_n=e_n|z) \approx P(E_n=e_n|z)$ for each error bit~$n$ and outputs 
$\hat{e}=(\hat e_1, \hat e_2, \dots, \hat e_N)$ such that $$\hat e_n = \argmax_{e_n\in\{0,1\}} \hat P(e_n|z).$$
{In addition, if $H\hat{e}=z$, then $\hat{e}$ is the decoder output.}
These marginal distributions can be calculated efficiently by \textit{message passing} on the Tanner graph.
If the Tanner graph has no cycles, then the BP algorithm will output a valid error vector~$\hat{e}$  with $H\hat{e}= z$ and the exact marginal distributions~\cite{Pea88,Wib96,MMC98,AM00,KFL01,YFW03}, i.e., $\hat P(e_n|z)= P(e_n|z)$ for $n=1,\dots,N$.
Even if there are cycles, the approximation is usually very good for a well-designed parity-check matrix~$H$ (e.g., no short cycles) \cite{Wib96,MMC98,AM00}.
The binary BP decoding algorithm is given in Algorithm~\ref{alg:CBP}~\cite{Mac99}. 
The order of message passing between the nodes is referred to as an \textit{updating schedule}.
Since the calculations at all the nodes in each step can be run in parallel,
message passing in this way is said to follow a \textit{parallel schedule}.
Consequently, Algorithm~\ref{alg:CBP} will be called \emph{parallel BP$_2$} in the following.

\begin{algorithm}[htbp] \caption{: Conventional binary BP decoding with a parallel schedule (parallel BP$_2$)} \label{alg:CBP}
\textbf{Input}: $H\in \{0,1\}^{M\times N}$, $z\in\{0,1\}^M$, and $\{(p_n^{(0)},p_n^{(1)})\}_{n=1}^N$.\\
{\bf Initialization.} For every variable node $n=1$ to $N$ and for every $m\in\sM(n)$, do:
	\begin{itemize}
	\item Let $q_{mn}^{(0)} = p_n^{(0)}$ and $q_{mn}^{(1)}=p_n^{(1)}$.
	\item Calculate 
	\begin{align}d_{mn}=q_{mn}^{(0)}-q_{mn}^{(1)} \label{eq:dmn}
	\end{align}
	 and pass it as the initial message $n\to m$.
	\end{itemize}
{\bf Horizontal Step.} For every check node $m=1$ to $M$ and  for every variable node $n\in\sN(m)$, compute
	\begin{align}
	\dlt_{mn} = (-1)^{z_m}\prod_{n'\in\sN(m)\setminus n} d_{mn'} \label{eq:delta_nm_BP2}
	\end{align}
	\begin{itemize}
	\item[] and pass it as the message $m\to n$.	
	\end{itemize}
{\bf Vertical Step.} For every variable node $n=1$ to $N$ and for every check node $m\in\sM(n)$, do:
	\begin{itemize}
	\item   Compute 
		\begin{align}
		r_{mn}^{(0)} &= (1+\dlt_{mn})/2, ~ r_{mn}^{(1)} = (1-\dlt_{mn})/2,\\
		q_{mn}^{(0)} &= a_{mn}\,p_n^{(0)}\prod_{m'\in\sM(n)\setminus m} r_{m'n}^{(0)}, \label{eq:qmn0}\\
		q_{mn}^{(1)} &= a_{mn}\,p_n^{(1)}\prod_{m'\in\sM(n)\setminus m} r_{m'n}^{(1)}, \label{eq:qmn1}
		\end{align}
		where $a_{mn}$ is a chosen scalar such that $q_{mn}^{(0)}+q_{mn}^{(1)}=1$.
	\item Update: $d_{mn} = q_{mn}^{(0)} - q_{mn}^{(1)}$ and pass it as the message $n\to m$.
	\end{itemize}
{\bf Hard Decision.} For every variable node $n=1$ to $N$, compute
	\begin{align}
	q_n^{(0)} &= p_n^{(0)}\prod_{m\in\sM(n)} r_{mn}^{(0)}, \label{eq:qn0} \\
	q_n^{(1)} &= p_n^{(1)}\prod_{m\in\sM(n)} r_{mn}^{(1)}.\label{eq:qn1}
	\end{align}
	\begin{itemize}
	\item[] Let $\hat e_n = 0$, if $q_n^{(0)}>q_n^{(1)}$, and $\hat e_n = 1$, otherwise.
	\end{itemize}

	\begin{itemize}
	\item Let $\hat{e} = (\hat e_1,\hat e_2,\dots,\hat e_N)$.
		\begin{itemize}
		\item  If $H\hat{e} = z$, 
		halt  and return ``SUCCESS'';
		\item otherwise, if a maximum number of iterations is reached, halt   and return ``FAIL'';
		\item otherwise, repeat from the horizontal step.
		\end{itemize}
	\end{itemize}
\end{algorithm}

Next we briefly explain parallel BP$_2$.  
 Let $p_n^{(0)}$ and $p_n^{(1)}$ be the probabilities  of $E_n$ being $0$ and $1$, respectively, for $n=1,\dots, N$,
 which are given by the underlying noisy channel. 
{Herein we assume a memoryless binary symmetric channel (BSC) with cross probability $\eps\in(0,0.5)$.}
Hence $p_n^{(0)}$ and $p_n^{(1)}$ are initialized as 
\begin{align*}
p_n^{(0)} &= P(E_n=0)=1-\eps \quad\text{and}\\
p_n^{(1)} &= P(E_n=1)=\eps.
\end{align*}
These channel parameters will be used in the generation of messages.

A message sent from variable node $n$ to check node $m$ will be denoted by message $n\to m$ for simplicity, and vice versa.
Let $\sN(m)$ denote the set of neighboring variable nodes of check node $m$
and let $\sM(n)$ denote the set of   neighboring check nodes of variable node $n$.
In BP, message $d_{n\to m}$ (defined~in~(\ref{eq:dmn})) will be passed from variable node $n$ to its neighboring check node $m$ 
and message $\dlt_{m\to n}$ (defined~in~(\ref{eq:delta_nm_BP2})) will be passed from check node $m$ to its neighboring variable node $n$.
{The messages $d_{n\to m}$ and $\dlt_{m\to n}$  will be denoted, respectively, by $d_{mn}$ and $\dlt_{mn}$ for simplicity, meaning that they are passed on the same edge associated with $H_{mn}$.}
Note that each of the passed messages $d_{mn}$ or $\delta_{mn}$ is a real number  {(representing the likelihood difference),} and they provide sufficient information for decoding.

 Let $q_n^{(0)}$ and $q_n^{(1)}$ be the likelihoods of $E_n$ being $0$ and~$1$, respectively, for $n=1,\dots, N$.
 These quantities will be updated as in (\ref{eq:qn0}) and (\ref{eq:qn1}) after a horizontal step and a vertical step and their sizes are used to estimate $E_n$.
The horizontal, vertical, and hard decision steps will be iterated until 
that an estimated error and the given syndrome vector are matched or a pre-defined maximum number of iterations is reached.

\begin{figure*}[hbp] \centerline{
		\subfigure[Initialization]{		
			\begin{tikzpicture}[node distance=1.3cm,>=stealth',bend angle=45,auto]

\tikzstyle{chk}=[rectangle,thick,draw=black!75,fill=black!20,minimum size=4mm]
\tikzstyle{var}=[circle,thick,draw=blue!75,fill=gray!20,minimum size=4mm,font=\footnotesize]
\tikzstyle{VAR}=[circle,thick,draw=blue!75,fill=blue!20,minimum size=5mm,font=\footnotesize]
\tikzstyle{fac}=[anchor=west,font=\footnotesize]

\node[var] (x3) at (0,0) {$E_3$};
\node[var] (x2) at (0,1) {$E_2$};
\node[var] (x1) at (0,2) {$E_1$};
\node[chk] (c1) at (1.5,2) {};
\node[chk] (c2) at (1.5,1) {};
\draw[->] (x1) -- (c1);
\draw[->] (x2) -- (c1);
\draw[->] (x1) -- (c2);
\draw[->] (x2) -- (c2);
\draw[->] (x3) -- (c2);

\end{tikzpicture}
			\label{fig:PBP_init}}
		\hfil
		\subfigure[Horizontal Step]{	
			\begin{tikzpicture}[node distance=1.3cm,>=stealth',bend angle=45,auto]

\tikzstyle{chk}=[rectangle,thick,draw=black!75,fill=black!20,minimum size=4mm]
\tikzstyle{var}=[circle,thick,draw=blue!75,fill=gray!20,minimum size=4mm,font=\footnotesize]
\tikzstyle{VAR}=[circle,thick,draw=blue!75,fill=blue!20,minimum size=5mm,font=\footnotesize]
\tikzstyle{fac}=[anchor=west,font=\footnotesize]

\node[var] (x3) at (0,0) {$E_3$};
\node[var] (x2) at (0,1) {$E_2$};
\node[var] (x1) at (0,2) {$E_1$};
\node[chk] (c1) at (1.5,2) {};
\node[chk] (c2) at (1.5,1) {};
\draw[<-] (x1) -- (c1);
\draw[<-] (x2) -- (c1);
\draw[<-] (x1) -- (c2);
\draw[<-] (x2) -- (c2);
\draw[<-] (x3) -- (c2);

\end{tikzpicture}
			\label{fig:PBP_hz}}
		\hfil
		\subfigure[Vertical Step]{		
			\begin{tikzpicture}[node distance=1.3cm,>=stealth',bend angle=45,auto]

\tikzstyle{chk}=[rectangle,thick,draw=black!75,fill=black!20,minimum size=4mm]
\tikzstyle{var}=[circle,thick,draw=blue!75,fill=gray!20,minimum size=4mm,font=\footnotesize]
\tikzstyle{VAR}=[circle,thick,draw=blue!75,fill=blue!20,minimum size=5mm,font=\footnotesize]
\tikzstyle{fac}=[anchor=west,font=\footnotesize]

\node[var] (x3) at (0,0) {$E_3$};
\node[var] (x2) at (0,1) {$E_2$};
\node[var] (x1) at (0,2) {$E_1$};
\node[chk] (c1) at (1.5,2) {};
\node[chk] (c2) at (1.5,1) {};
\draw[->] (x1) -- (c1);
\draw[->] (x2) -- (c1);
\draw[->] (x1) -- (c2);
\draw[->] (x2) -- (c2);
\draw[->] (x3) -- (c2);

\end{tikzpicture}	
			\label{fig:PBP_vt}}
	}
	\caption{The order of message passing in Algorithm~\ref{alg:CBP} (parallel BP$_2$) for the example in Fig.~\ref{fig:H2x3}.
		(a) Initialization: For every $n$ and for $m\in\sM(n)$, message $d_{mn}$ is initialized to $p_n^{(0)}-p_n^{(1)}$ and then passed from variable node $n$ to check node $m$.
		(b) Horizontal Step:  For every $m$ and for $n\in\sN(m)$, message $\dlt_{mn}$ is computed and then passed 
			from check node $m$ to variable node $n$.
		(c) Vertical Step: For every $n$ and for $m\in\sM(n)$, message $d_{mn}$ is computed and then passed 
			from variable node $n$ to check node $m$.
		Then  (b) and (c) are repeated in the following iterations.
	}\label{fig:PBP2}
\end{figure*}
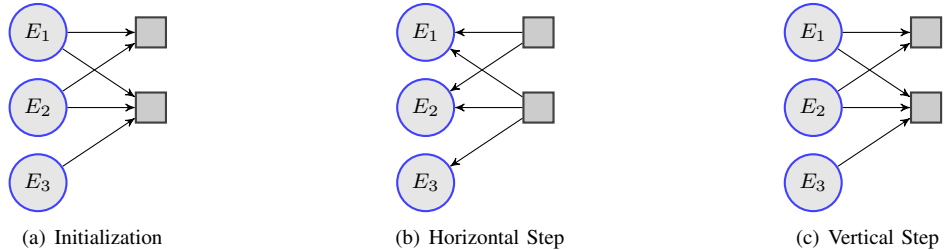

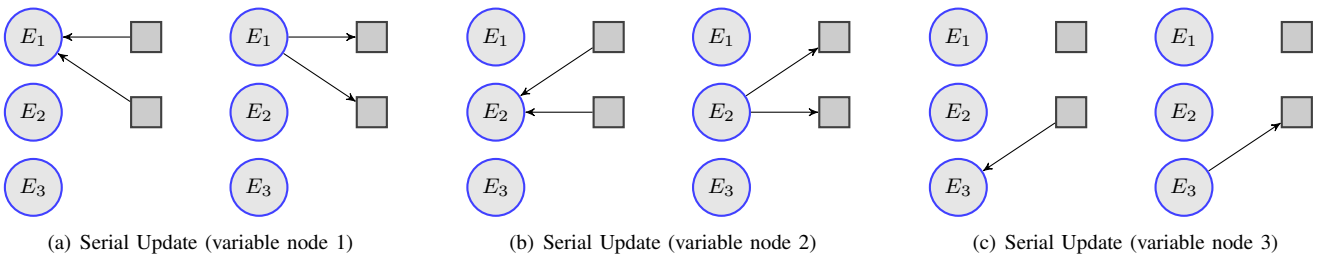
\begin{figure*}[hbp] \centerline{
		\subfigure[Serial Update (variable node 1)]{	
			\begin{tikzpicture}[node distance=1.3cm,>=stealth',bend angle=45,auto]

\tikzstyle{chk}=[rectangle,thick,draw=black!75,fill=black!20,minimum size=4mm]
\tikzstyle{var}=[circle,thick,draw=blue!75,fill=gray!20,minimum size=4mm,font=\footnotesize]
\tikzstyle{VAR}=[circle,thick,draw=blue!75,fill=blue!20,minimum size=5mm,font=\footnotesize]
\tikzstyle{fac}=[anchor=west,font=\footnotesize]

\node[var] (x1) at (0,5.5) {$E_1$};
\node[var] (x2) at (0,4.5) {$E_2$};
\node[var] (x3) at (0,3.5) {$E_3$};
\node[chk] (c1) at (1.5,5.5) {};
\node[chk] (c2) at (1.5,4.5) {};
\draw[<-] (x1) -- (c1);
\draw[<-] (x1) -- (c2);

\node[var] (x1) at (3,5.5) {$E_1$};
\node[var] (x2) at (3,4.5) {$E_2$};
\node[var] (x3) at (3,3.5) {$E_3$};
\node[chk] (c1) at (4.5,5.5) {};
\node[chk] (c2) at (4.5,4.5) {};
\draw[->] (x1) -- (c1);
\draw[->] (x1) -- (c2);

\end{tikzpicture}
			\label{fig:SBP_var1}}~~~~
		\hfil
		\subfigure[Serial Update (variable node 2)]{	
			\begin{tikzpicture}[node distance=1.3cm,>=stealth',bend angle=45,auto]

\tikzstyle{chk}=[rectangle,thick,draw=black!75,fill=black!20,minimum size=4mm]
\tikzstyle{var}=[circle,thick,draw=blue!75,fill=gray!20,minimum size=4mm,font=\footnotesize]
\tikzstyle{VAR}=[circle,thick,draw=blue!75,fill=blue!20,minimum size=5mm,font=\footnotesize]
\tikzstyle{fac}=[anchor=west,font=\footnotesize]

\node[var] (x1) at (0,5.5) {$E_1$};
\node[var] (x2) at (0,4.5) {$E_2$};
\node[var] (x3) at (0,3.5) {$E_3$};
\node[chk] (c1) at (1.5,5.5) {};
\node[chk] (c2) at (1.5,4.5) {};
\draw[<-] (x2) -- (c1);
\draw[<-] (x2) -- (c2);

\node[var] (x1) at (3,5.5) {$E_1$};
\node[var] (x2) at (3,4.5) {$E_2$};
\node[var] (x3) at (3,3.5) {$E_3$};
\node[chk] (c1) at (4.5,5.5) {};
\node[chk] (c2) at (4.5,4.5) {};
\draw[->] (x2) -- (c1);
\draw[->] (x2) -- (c2);

\end{tikzpicture}
			\label{fig:SBP_var2}}~~~~
		\hfil
		\subfigure[Serial Update (variable node 3)]{	
			\begin{tikzpicture}[node distance=1.3cm,>=stealth',bend angle=45,auto]

\tikzstyle{chk}=[rectangle,thick,draw=black!75,fill=black!20,minimum size=4mm]
\tikzstyle{var}=[circle,thick,draw=blue!75,fill=gray!20,minimum size=4mm,font=\footnotesize]
\tikzstyle{VAR}=[circle,thick,draw=blue!75,fill=blue!20,minimum size=5mm,font=\footnotesize]
\tikzstyle{fac}=[anchor=west,font=\footnotesize]

\node[var] (x1) at (0,5.5) {$E_1$};
\node[var] (x2) at (0,4.5) {$E_2$};
\node[var] (x3) at (0,3.5) {$E_3$};
\node[chk] (c1) at (1.5,5.5) {};
\node[chk] (c2) at (1.5,4.5) {};
\draw[<-] (x3) -- (c2);

\node[var] (x1) at (3,5.5) {$E_1$};
\node[var] (x2) at (3,4.5) {$E_2$};
\node[var] (x3) at (3,3.5) {$E_3$};
\node[chk] (c1) at (4.5,5.5) {};
\node[chk] (c2) at (4.5,4.5) {};
\draw[->] (x3) -- (c2);

\end{tikzpicture}
			\label{fig:SBP_var3}}
	}
	\caption{The order of message passing in Algorithm~\ref{alg:SBP} (serial BP$_2$) for the example in Fig.~\ref{fig:H2x3}.
		The initialization procedure is the same as in Fig.~\ref{fig:PBP2}\,(a).
		In Serial Update: 
		(a) Variable node 1 receives $\dlt_{11}$ and $\dlt_{21}$, updates $d_{11}$ and $d_{21}$ and sends them to the two check nodes, respectively. (b) and (c) are similar to (a) but are with respect to variable nodes 2 and 3, respectively.
		Then Serial Update (a), (b), and (c) are iterated.	
	}\label{fig:SBP2}
\end{figure*}

We illustrate how parallel BP$_2$ works with a  simple but essential example, which can be extended to the quantum case later in Sec.~\ref{sec:QuanBP}.
Consider again the parity-check matrix
$H=\left[\begin{smallmatrix}1 &1 &0\\ 1 &1 &1\end{smallmatrix}\right]$
with Tanner graph given in Fig.~\ref{fig:H2x3}.
Given error syndrome $(z_1,z_2)\in\{0,1\}^2$, $H$ imposes the two parity-check constraints: 
\begin{itemize}
\item The first parity check $H_1$: $E_1+E_2 = z_1$.
\item The second parity check $H_2$: $E_1+E_2+E_3 = z_2$.
\end{itemize}
For convenience, here we analyze the algorithm in terms of likelihood ratio of the first variable bit $E_1$, denoted by $LR_1$,
and it is similar for the other error bits. 
When $LR_1$ is larger than $1$, $E_1$ is more likely to be $0$ than $1$.
Initially, $LR_1$ is $\frac{p_1^{(0)}}{p_1^{(1)}}$ from the channel parameters.
Then it is updated at each iteration of message passing.
After the first iteration, we have
\begin{align*}
LR_1&= \frac{q_1^{(0)}}{q_1^{(1)}} ~=~ \frac{p_1^{(0)}}{p_1^{(1)}}\times \frac{r_{11}^{(0)}}{r_{11}^{(1)}} \times \frac{r_{21}^{(0)}}{r_{21}^{(1)}} \\
	&= \frac{p_1^{(0)}}{p_1^{(1)}}\times \left(\frac{p_2^{(0)}}{p_2^{(1)}}\right)^{(-1)^{z_1}} \times \left(\frac{p_2^{(0)}p_3^{(0)}+p_2^{(1)}p_3^{(1)}}{p_2^{(0)}p_3^{(1)}+p_2^{(1)}p_3^{(0)}}\right)^{(-1)^{z_2}}_.
\end{align*}
The second and the third terms are contributed by the parity checks $H_1$ and $H_2$, respectively.
For example, if the second syndrome bit is $z_2=0$, we have  $E_1=E_2+E_3$. If $E_1=0$, then $(E_2,E_3)=(0,0)$ or $(1,1)$;
if $E_1=1$, then $(E_2,E_3)=(0,1)$ or $(1,0)$. Thus we have a belief contribution of $\frac{p_2^{(0)}p_3^{(0)}+p_2^{(1)}p_3^{(1)}}{p_2^{(0)}p_3^{(1)}+p_2^{(1)}p_3^{(0)}}$   from $H_2$.
Passing the single-valued messages $\delta_{mn}$ calculated in (\ref{eq:delta_nm_BP2}) is sufficient to complete these belief updates. 
In the meanwhile messages $d_{mn}$ are updated for the next iteration. 
Therefore, the $\delta$-rule works well for BP$_2$.

In parallel BP$_2$, the messages are updated according to a parallel schedule. 
In general, a shuffled or serial update schedule may have some benefits~\cite{ZF05,SLG07}.
Algorithm~\ref{alg:SBP}, referred to as \mbox{\emph{serial BP$_2$}}, defines a BP decoding algorithm according to  a serial schedule. 
A serial update can be done along the variable nodes or along the check nodes {(with similar performance)} \cite{SLG07} and in Algorithm~\ref{alg:SBP} the update is along the variable nodes.

\begin{algorithm}[htbp] \caption{: Binary BP decoding according to a serial schedule along the variable nodes (serial BP$_2$)}\label{alg:SBP}
\textbf{Input}: $H\in \{0,1\}^{M\times N}$, $z\in\{0,1\}^M$, and $\{(p_n^{(0)},p_n^{(1)})\}_{n=1}^N$.\\
{\bf Initialization.} Do the same as in Algorithm~\ref{alg:CBP}. \\
{\bf Serial Update.} For each variable node $n=1$ to $N$, do:
	\begin{itemize}
	\item For each check node $m\in\sM(n)$,  compute\\  
		$\dlt_{mn}= (-1)^{z_m}\prod_{n'\in\sN(m)\setminus n} d_{mn'}$ \\
	and pass it as the \mbox{message $m\to n$}.
	\item 
		For each $m\in\sM(n)$, compute\\
		$r_{mn}^{(0)} = (1+\dlt_{mn})/2, ~ r_{mn}^{(1)} = (1-\dlt_{mn})/2,$ \\
		$q_{mn}^{(0)} = a_{mn}\,p_n^{(0)}\prod_{m'\in\sM(n)\setminus m} r_{m'n}^{(0)}$, \\
		$q_{mn}^{(1)} = a_{mn}\,p_n^{(1)}\prod_{m'\in\sM(n)\setminus m} r_{m'n}^{(1)}$, \\
		where $a_{mn}$ is a chosen scalar such that $q_{mn}^{(0)}+q_{mn}^{(1)}=1$.
	\item Update: $d_{mn} =q_{mn}^{(0)}-q_{mn}^{(1)}$ and pass it as the message $n\to m$.
	\end{itemize}
{\bf Hard Decision.} 
	\begin{itemize}
	\item Do the same as in Algorithm~\ref{alg:CBP}, except that ``repeat from the horizontal step'' is replaced by 
	``repeat from the serial update step''.
	\end{itemize}
\end{algorithm}

%

{Despite of the different schedules, serial BP$_2$ and parallel BP$_2$ have to update the same number of messages ($d_{mn}$ and $\delta_{mn}$) and thus have the same computational complexity in an iteration.
	While parallel BP$_2$ achieves a full parallelism in a horizontal step and a vertical step, 
	serial BP$_2$ tries to utilize the most updated $d_{mn}$ to accelerate the convergence 
	at the cost of parallelism. (A partial parallelism  is still possible by a careful design \cite{ZF05,SLG07}.)  
For clarity, we use the example in Fig.~\ref{fig:H2x3} to show how the parallel and serial schedules work in Figures~\ref{fig:PBP2}
and~\ref{fig:SBP2}.
The main advantage of serial BP$_2$ is that it converges in roughly half the number of iterations, compared to parallel BP$_2$, to achieve the same accuracy \cite{ZF05,SLG07}.
More precisely,
the advantage/disadvantage of the two schedules can be understood as follows. If  full parallelism is possible with sufficient hardware resources, 
each iteration of the parallel~BP$_2$ takes less time. Otherwise, the parallel BP$_2$ and serial BP$_2$  may run roughly the same number of iterations in a fixed time, but the convergence behavior of serial BP$_2$ is usually better.

For illustration,  we consider the $[13298,3296]$ code defined in \cite{Mac99}. The performance of parallel BP$_2$ is shown in Fig.~\ref{fig:BP20} with respect to various maximum numbers of iterations.
The maximum numbers of iterations for the curves from the left-hand-side to the right-hand-side are 10, 15, 20, 25, 30, 40, 50, and 100, respectively. An error bar between two crosses shows a $95\%$ confidence interval.  Similarly, the performance of serial BP$_2$ is shown in Fig.~\ref{fig:BP24}.
A successful decoding is counted when the decoder converges to the actual error. 
Otherwise, a block error occurs, and we could have either a \emph{detected error} if the syndrome  is not matched after a maximum number of iterations is reached or an \emph{undetected error} if the syndrome is falsely matched because the decoder converges to a wrong solution. 
We have the same observation by MacKay~\cite{Mac99} that all the observed block errors are detected errors for this code.
For a data point with a largest block error rate $\le 10^{-4}$, we calculate the average number of iterations  by including all the successful and block-errored cases. The average numbers of iterations for a maximum of 15, 30, and 100 iterations are given in Table~\ref{tbl:avgiter}.
%
%
%
In this case, both schedules converge well if the computational power is enough. 
(However, this may not be the case for BP decoding of quantum codes (due to the many short cycles)
and an interesting example will be shown later in Fig.~\ref{fig:Y4}.)

Unless otherwise stated, the simulations in later sections will follow the same criteria as above.

	\begin{figure}[tbp]
	\centering \includegraphics[width=0.5\textwidth]{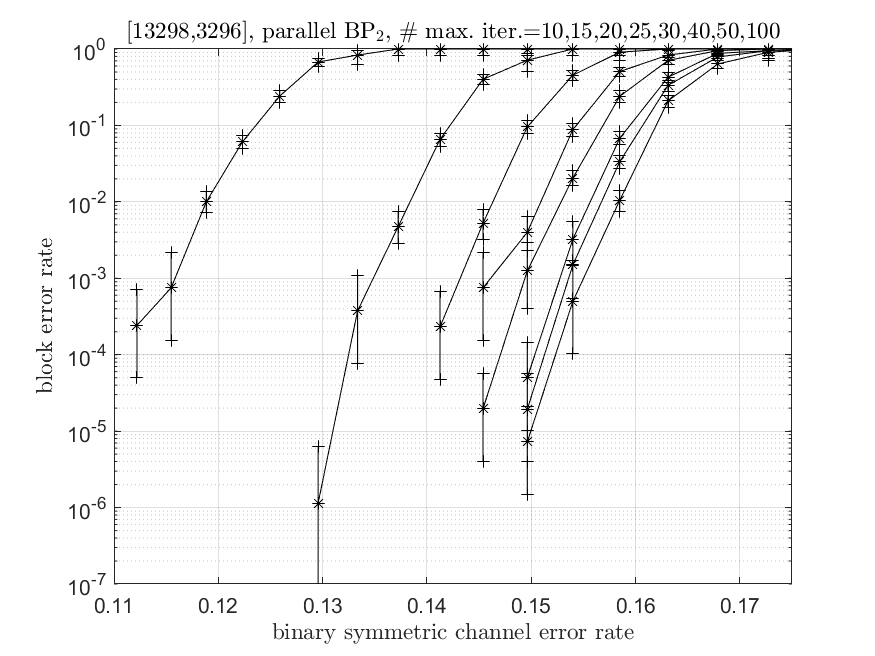}
	\caption{Parallel BP$_2$ decoding in different maximum numbers of iterations} \label{fig:BP20}	\vspace*{\floatsep}
	\centering \includegraphics[width=0.5\textwidth]{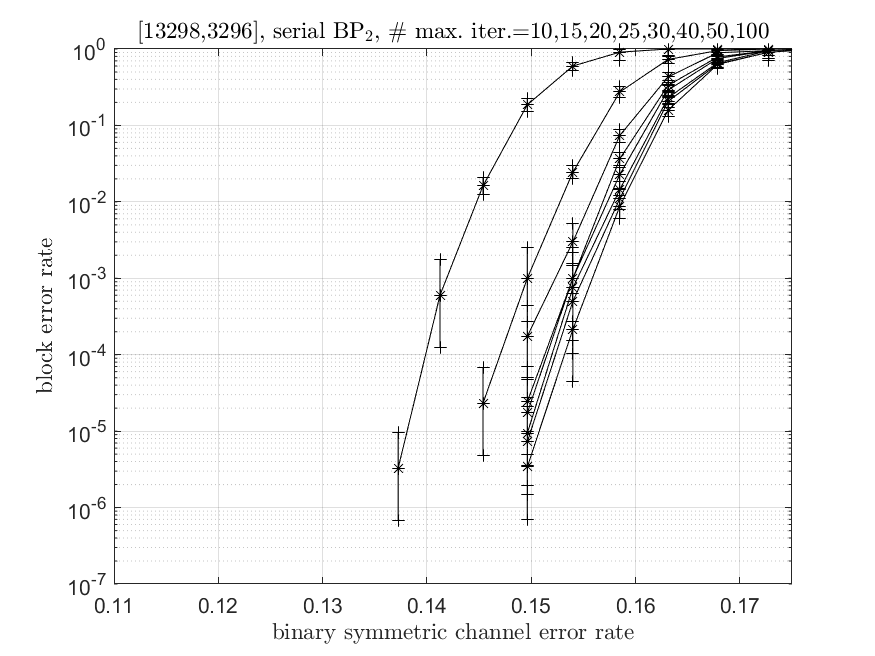}
	\caption{Serial BP$_2$ decoding in different maximum numbers of iterations} \label{fig:BP24}
	\end{figure}

\begin{table}[htbp] \caption{
The  average number of   iterations (Iter$_\text{avg}$) to decode the $[13298,3296]$ code 
for a maximum number of 15,30, and 100 iterations (Iter$_\text{max}$). The target block error rate is $\leq 10^{-4}$. 
} \label{tbl:avgiter} \centering
	$\begin{array}{|c|c|c|c|}
	\hline
	\text{Iter$_\text{avg}$} 
							& \text{Iter}_\text{max}=15	& \text{Iter}_\text{max}=30	& \text{Iter}_\text{max}=100 \\ \hline
	\text{Parallel BP$_2$}	& 10.72						& 15.16						& 17.3			\\
	\text{Serial BP$_2$}	& 8.37						& 9.43						& 9.44			\\
	\hline
	\end{array}$
\end{table}

\section{Quaternary BP Decoding for Quantum Codes} \label{sec:QuanBP}

\subsection{Tanner graph and belief propagation for quantum stabilizer codes}

We focus on binary stabilizer codes for quantum information in qubits and consider error operators that are tensor product of Pauli matrices 
\mbox{$\left\{	I=\left[\begin{smallmatrix}1 &0\\0&1\end{smallmatrix}\right], 
				X=\left[\begin{smallmatrix}0 &1\\1&0\end{smallmatrix}\right], 
				Y=\left[\begin{smallmatrix}0 &-i\\i&0\end{smallmatrix}\right], 
				Z=\left[\begin{smallmatrix}1 &0\\0&-1\end{smallmatrix}\right]\right\}$ \cite{Shor95,GotPhD,CS96,Steane96}. 
}
Specifically, we consider an independent \textit{depolarizing} channel with rate $\epsilon$ such that the probability of Pauli errors $\{I,X,Y,Z\}$ is 
$$	{\mbf p} = (p^I,p^X,p^Y,p^Z)= (1-\eps,\eps/3,\eps/3,\eps/3).	$$
The \textit{weight} of an $N$-fold Pauli operator is the number of its nonidentity entries.
A low-weight Pauli error occurs more likely than a high-weight error in a depolarizing channel.

Since Pauli matrices either commute or anticommute with each other, we can define an inner product  $\langle \cdot, \cdot \rangle:\{I,X,Y,Z\}\times \{I,X,Y,Z\} \rightarrow \{0,1\}$ for Pauli matrices as in Table~\ref{tbl:bES} such that for $E_1,E_2\in\{I,X,Y,Z\}$,
$\langle E_1,E_2\rangle=1$ if they anticommute with each other, and $\langle E_1,E_2\rangle=0$ if they commute with each other.
This inner product can be naturally extended to an inner product   for $N$-fold Pauli operators 
$\langle \cdot,\cdot\rangle: \{I,X,Y,Z\}^{\otimes N}\times \{I,X,Y,Z\}^{\otimes N} \rightarrow \{0,1\}$, defined by 
\begin{equation}\label{inner_prod}
	\langle E,F \rangle = \sum_{n=1}^N\langle E_n,F_n\rangle  \mod 2,
\end{equation}
where $E=E_1\otimes E_2\otimes \cdots \otimes E_N$, $F=F_1\otimes F_2\otimes \cdots \otimes F_N\in\{I,X,Y,Z\}^{\otimes N}$.
Note that we use the same notation for inner product without ambiguity as the value indicates whether two $N$-fold Pauli operators commute with each other or not.
From now on the tensor product $\otimes$ will be omitted.

\begin{table}[htbp] \caption{Commutation Relations of Pauli Operators (0:\,commute,~1:\,anticommute)} \label{tbl:bES} \centering
	$\begin{array}{|c|c|c|c|c|}
	\hline
	\langle E_n,F_n\rangle 
	& F_n=I	& F_n=X	& F_n=Y	& F_n=Z	\\ \hline
	E_n=I	& 0			& 0			& 0			& 0			\\
	E_n=X	& 0			& 0			& 1			& 1			\\
	E_n=Y	& 0			& 1			& 0			& 1			\\
	E_n=Z	& 0			& 1			& 1			& 0			\\
	\hline
	\end{array}$
\end{table}

An $[[N,K]]$ quantum stabilizer code is a $2^K$-dimensional subspace of $\mathbb{C}^{2^N}$.
It can be defined by a stabilizer check matrix $S\in\{I,X,Y,Z\}^{M\times N}$ (not necessarily of full rank) with $M\geq N-K$. 
Each row $S_m$ of $S$ corresponds to an $N$-fold Pauli operator that stabilizes the code space, i.e., the code space is contained in its $(+1)$-eigenspace.
The matrix $S$ is self-orthogonal with respect to the inner product \eq{inner_prod}, i.e., 
$\langle S_m,S_{m'}\rangle=0$ for any two rows $S_m$ and $S_{m'}$ of $S$.
In fact, the code space is the joint-($+1$) eigenspace of the rows of  $S$, and
the vectors in the rowspace of $S$ are  called \emph{stabilizers} \cite{NC00}.

We assume that quantum information is initially  encoded by a noiseless stabilizer circuit~\cite{GotPhD,KL19}   
and then the encoded state suffers depolarizing errors. 
Therefore, we may assume that the encoded state is corrupted by an unknown \mbox{$N$-qubit}  error operator $E\in\{I,X,Y,Z\}^N$ with corresponding probability.

To do error correction,  the stabilizers   $\{S_m: m=1,2,\dots,M \}$  are measured to determine the \emph{binary error syndrome} $z = (z_1,z_2,\dots,z_M)\in\{0,1\}^M$, 
where 
\begin{align}
z_m= \langle E,S_m \rangle \in\{0,1\}. \label{eq:syndrome_bit}
\end{align}
Given $S$ and $z$, a decoder has to estimate an error $\hat E\in\{I,X,Y,Z\}^N$ such that $\hat E$ is equivalent to $E$, up to a stabilizer,  
with a probability as high as possible. 
Note that the solution is not unique due to the degeneracy of the quantum code~\cite{PC08,KL13_20}.

A Tanner graph corresponding to the $M\times N$ quantum stabilizer check matrix $S$ can be similarly defined as in the classical case: it is a bipartite graph consisting of $N$ variable nodes and $M$ check nodes and it has an edge connecting check node $m$ and variable node $n$ if $S_{mn}\neq I$.
However, there are three types of edges corresponding to $X,Y,Z$, respectively.
A stabilizer $S_m$ defines a relation as in (\ref{eq:syndrome_bit}). 
 An example of $S=\begin{bmatrix} X&Y&I\\Z&Z&Y\end{bmatrix}$ is shown in Fig.~\ref{fig:S2x3}.
Thus the quantum decoding problem can be handled by a quaternary BP (denoted by BP$_4$) on the Tanner graph. 
\begin{figure}[htbp] \centering ~~~~ \begin{tikzpicture}[node distance=1.3cm,>=stealth',bend angle=45,auto]

\tikzstyle{chk}=[rectangle,thick,draw=black!75,fill=black!20,minimum size=4mm]
\tikzstyle{var}=[circle,thick,draw=blue!75,fill=gray!20,minimum size=4mm,font=\footnotesize]
\tikzstyle{VAR}=[circle,thick,draw=blue!75,fill=blue!20,minimum size=5mm,font=\footnotesize]
\tikzstyle{fac}=[anchor=west,font=\footnotesize]

\node[var] (x3) at (0,0) {$E_3$};
\node[var] (x2) at (0,1) {$E_2$};
\node[var] (x1) at (0,2) {$E_1$};
\node[chk] (c1) at (1.5,2) {};
\node[chk] (c2) at (1.5,1) {};

\draw[thick] (x1) -- (c1);
\draw[thick,dashed] (x2) -- (c1);
\draw[thick,densely dotted] (x1) -- (c2) -- (x2);
\draw[thick,dashed] (x3) -- (c2);


\node[fac] [right of=c1,xshift=14mm] {$(S_1):~\langle E_1,X\rangle+\langle E_2,Y\rangle = z_1$};
\node[fac] [right of=c2,xshift=21mm] {$(S_2):~\langle E_1,Z\rangle+\langle E_2,Z\rangle+\langle E_3,Y\rangle = z_2$};

\node[fac] (Xl) [right of=x3,xshift=5mm,yshift=7] {};
\node[fac] (Xr) [right of=x3,xshift=20mm,yshift=7] {$X$};
\draw[thick] (Xl) -- (Xr);
\node[fac] (Yl) [right of=x3,xshift=5mm,yshift=0] {};
\node[fac] (Yr) [right of=x3,xshift=20mm,yshift=0] {$Y$};
\draw[thick,dashed] (Yl) -- (Yr);
\node[fac] (Zl) [right of=x3,xshift=5mm,yshift=-7] {};
\node[fac] (Zr) [right of=x3,xshift=20mm,yshift=-7] {$Z$};
\draw[thick,densely dotted] (Zl) -- (Zr);

\end{tikzpicture}
	\caption{The Tanner graph of   $S=\left[{X\atop Z}{Y\atop Z}{I\atop Y}\right]$.} \label{fig:S2x3}
\end{figure}

A conventional BP$_4$ for decoding binary stabilizer codes is done as follows \cite{PC08}. 
Initially, the channel parameters are ${\mbf p}_n= (p_n^I,p_n^X,p_n^Y,p_n^Z)$ for $n=1,\dots,N$, where
\beqs{	p_n^I ~&\,=P(E_n=I ) =\, 1-\eps, \quad \text{and} \\
	 	p_n^W  &\,=P(E_n=W ) =\, \eps/3, ~\quad \text{for $W\in\{X,Y,Z\}$}.
}
In the initialization step, 
at every variable node~$n$, pass the message
$\mbf{q}_{mn}=(q_{mn}^I,q_{mn}^X,q_{mn}^Y,q_{mn}^Z)$ $= {\mbf p}_n$   to every neighboring  check node $m\in\sM(n)$. 
As in \cite{Wib96}, let $E|_{\sN(m)}$ be the restriction of  $E=E_1E_2\cdots E_N$ to the coordinates in~$\sN(m)$. 
Note that $\langle E,S_m \rangle = \langle E|_{\sN(m)},S_m|_{\sN(m)} \rangle$ for any $E$ and $S_m$ since only the components of $S_m$ corresponding to $\sN(m)$ are not the identity $I$.\,\footnote{
	For example, if $S_m=IXZ$ and $E=E_1E_2E_3$, then\\ \mbox{$E|_{\sN(m)}=(E_1E_2E_3)|_{\sN(m)} = E_2E_3$} and $S_m|_{\sN(m)}= XZ$. Then\\
	$\langle E,S_m \rangle = \langle E_1E_2E_3, IXZ \rangle = \left\langle E_2E_3, XZ \right\rangle = \langle E|_{\sN(m)},S_m|_{\sN(m)} \rangle.$
	}
Then according to a parallel update schedule, the check nodes and variable nodes work as follows. 
\begin{itemize} 
\item Horizontal Step:
At check node $m$, compute $\mbf{r}_{mn}=(r_{mn}^I,r_{mn}^X,r_{mn}^Y,r_{mn}^Z)$ and pass $\mbf{r}_{mn}$ as the message \mbox{$m\to n$}
for every \mbox{$n\in\sN(m)$}, where
\begin{align}
r_{mn}^W 
	&= \sum_{E|_{\sN(m)}:~E_n=W, \atop \left\langle E|_{\sN(m)},S_m|_{\sN(m)} \right\rangle=z_m} \ps{ \prod_{n'\in\sN(m)\setminus n}q_{mn'}^{E_{n'}} }	\label{rml}
\end{align}
for $W\in\{I,X,Y,Z\}$.

\item Vertical Step:
At variable node $n$, compute $\mbf{q}_{mn}=(q_{mn}^I,q_{mn}^X,q_{mn}^Y,q_{mn}^Z)$ and pass $\mbf{q}_{mn}$
as the message \mbox{$n\to m$} for every $m\in\sM(n)$, where
\beql{qml}{ q_{mn}^W = a_{mn}\, p_{n}^W \prod_{m'\in\sM(n)\setminus m} r_{m'n}^W 
}
for $W\in\{I,X,Y,Z\}$ and $a_{mn}$ is a chosen scalar such that $q_{mn}^I+q_{mn}^X+q_{mn}^Y+q_{mn}^Z=1$.
\end{itemize} 
At variable node $n$, it also computes $q_n^W = p_{n}^W \prod_{m\in\sM(n)} r_{mn}^W $ for $W\in\{I,X,Y,Z\}$ (where normalization is not necessary).
A hard decision is made by $\hat E_n = \argmax_{W\in\{I,X,Y,Z\}} q_n^{W}$.
The horizontal step and the vertical step are iterated until an estimated error $\hat{E}=\hat{E}_1\cdots \hat{E}_N$ is valid or a maximum number of iterations is reached.

In general, BP$_4$  requires  higher computing complexity than BP$_2$ as mentioned in the introduction. Moreover, the $\delta$-rule used in Algorithm~\ref{alg:CBP} cannot be  applied to BP$_4$ for classical quaternary codes.
While the variable-node computation~\eq{qml} is relatively straightforward,   
the check-node computation~\eq{rml} seems to have a large room for further simplification. 
We will show that this computation can also be simplified by a $\delta$-rule in the following subsection and then we can design a BP algorithm for stabilizer codes so that the passed messages are single-valued as in the case of BP$_2$ using the $\delta$-rule. Thus the complexity of our BP decoding algorithm for stabilizer codes is significantly reduced, compared to the conventional BP$_4$ for stabilizer codes.

\subsection{Refined belief propagation decoding  of stabilizer codes}
An important observation is that the error syndrome of a binary stabilizer code (\ref{eq:syndrome_bit}) is binary. 
Given  the value of $\langle E_n, S_{mn}\rangle$ for (unknown) $E_n\in\{I,X,Y,Z\}$ and some $S_{mn}\in\{X,Y,Z\}$, we will know that $E_n$ commutes or anticommutes with $S_{mn}$, i.e.,  either $E_n\in \{I, S_{mn}\}$ or $E_n\in\{X,Y,Z\}\setminus S_{mn}$.
Consequently, the passed message should indicate more likely whether $E_n\in \{I, S_{mn}\}$ or $E_n\in\{X,Y,Z\}\setminus S_{mn}$.
For a variable node, say variable node~1, and its neighboring check node $m$, we know that from~(\ref{eq:syndrome_bit}) 
\begin{align*}
\langle E_1, S_{m1}\rangle &= z_m+\sum_{n=2}^N \langle E_n, S_{mn}\rangle \mod 2.  
\end{align*}
In other words, the message from a neighboring check will tell us more likely whether the error $E_1$ commutes or anticommutes with $S_{m1}$. 
This suggests that a BP decoding of stabilizer codes with single-valued messages is possible and we provide such an algorithm in Algorithm~\ref{alg:QBP}, which is referred to as \textit{parallel BP$_4$}. We simplify a notation $r_{mn}^{(\langle W,S_{mn}\rangle)}$ as $r_{mn}^{\langle W,S_{mn}\rangle}$.

\begin{algorithm}[htbp] 
	\caption{:  $\dlt$-rule based quaternary BP decoding for binary stabilizer codes  	with a parallel schedule (parallel BP$_4$)} \label{alg:QBP}
	\textbf{Input}:  $S \in\{I,X,Y,Z\}^{M\times N}$, $z \in\{0,1\}^M$, and initial $\{(p_n^I, p_n^X, p_n^Y, p_n^Z)\}_{n=1}^N$.\\
	{\bf Initialization.}  For every variable node $n=1$ to $N$ and for every $m\in\sM(n)$, do: 
	\begin{itemize} 
		\item Let $q_{mn}^W = p_n^W$ for $W\in\{I,X,Y,Z\}$.
		\item Let $q_{mn}^{(0)} = q_{mn}^I+q_{mn}^{S_{mn}}$ and $q_{mn}^{(1)} = 1 - q_{mn}^{(0)}$. Calculate 
		\begin{equation}
		d_{mn}=q_{mn}^{(0)}-q_{mn}^{(1)} \label{eq:dmn_BP4}
		\end{equation}
		and pass it as the initial message $n\to m$.
	\end{itemize}
	
	{\bf Horizontal Step.} For every check node $m=1$ to $M$ and for   every $n\in\sN(m)$, compute 
	\begin{align}
	\dlt_{mn} = (-1)^{z_m}\prod_{n'\in\sN(m)\setminus n} d_{mn'}   \label{eq:delta_mn_BP4}
	\end{align}
	\begin{itemize}
	\item[] and pass it as the message $m\to n$. 
	\end{itemize}
	
	{\bf Vertical Step.} For every variable node $n=1$ to $N$ and for every $m\in\sM(n)$, do: 
	\begin{itemize}
		\item Compute 
		\begin{align}
		r_{mn}^{(0)} &= (1+\dlt_{mn})/2, ~ r_{mn}^{(1)} = (1-\dlt_{mn})/2,\\
		q_{mn}^I &= p_n^I\prod_{m'\in\sM(n)\setminus m} r_{m'n}^{(0)}, \label{eq:qmnI} \\
		q_{mn}^W &= p_n^W\prod_{m'\in\sM(n)\setminus m} r_{m'n}^{\langle W, S_{m'n}\rangle }, \mbox{ for } W\in\{X,Y,Z\}. \label{eq:qmnW}
		\end{align}
		\item Let $q_{mn}^{(0)} = a_{mn}(q_{mn}^I + q_{mn}^{S_{mn}})$\\ and $q_{mn}^{(1)} = a_{mn}(\sum_{W'\in\{X,Y,Z\}\setminus S_{mn}}q_{mn}^{W'})$,\\
		where  $a_{mn}$ is a chosen scalar such that $q_{mn}^{(0)}+q_{mn}^{(1)}=1$.
		\item Update: $d_{mn} = q_{mn}^{(0)} - q_{mn}^{(1)}$ and pass it as  the message $n\to m$.
	\end{itemize}
	
	{\bf Hard Decision.} 
						For every variable node $n=1$ to $N$, compute
		\begin{align}
		\qquad  q_n^I &= p_n^I\prod_{m\in\sM(n)} r_{mn}^{(0)}  \label{eq:qnI}\\
				q_n^W &= p_n^W\prod_{m\in\sM(n)} r_{mn}^{\langle W,S_{mn}\rangle}, \mbox{ for } W\in\{X,Y,Z\}. \label{eq:qnW}
		\end{align} 
		\begin{itemize}
		\item[]	Let $\hat E_n = \argmax_{W\in\{I,X,Y,Z\}} q_n^{W}.$
		\end{itemize}

	\begin{itemize}
		\item Let $\hat E = \hat E_1\hat E_2\cdots\hat E_N$. 
		\begin{itemize}
			\item If $\langle \hat E, S_m \rangle = z_m$ for $m=1,2,\dots,M$, halt   and return ``SUCCESS'';
			\item otherwise, if a maximum number of iterations is reached, halt   and return ``FAIL'';
			\item otherwise, repeat from the horizontal step.
		\end{itemize}
	\end{itemize}
\end{algorithm}

\begin{algorithm}[htbp] \caption{: BP$_4$ decoding of binary stabilizer codes with a serial schedule along the variable nodes (serial BP$_4$)} \label{alg:SQBP}
\textbf{Input:} $S \in\{I,X,Y,Z\}^{M\times N}$, $z \in\{0,1\}^M$, and initial $\{(p_n^I, p_n^X, p_n^Y, p_n^Z)\}_{n=1}^N$.\\
The remaining steps parallel those in Algorithm~\ref{alg:SBP} but with calculations replaced by those correspondences in
 Algorithm~\ref{alg:QBP}. (Details   are omitted.)
\end{algorithm}

It can be shown that Algorithm~\ref{alg:QBP} has exactly the same output as the conventional BP$_4$ outlined in the previous subsection but with an improved complexity similar to BP$_2$.
(The verification is straightforward and omitted here.)
In particular, Algorithm~\ref{alg:QBP} has a check-node efficiency 16-fold improved from   BP in GF(4)  as mentioned in Sec.~\ref{sec:Intro}.

Comparing Algorithms~\ref{alg:CBP} and \ref{alg:QBP},  one can find that (\ref{eq:dmn_BP4})--(\ref{eq:qnW}) parallel
(\ref{eq:dmn})--(\ref{eq:qn1}), respectively.
However, it is highly nontrivial to obtain these expressions for Algorithm~\ref{alg:QBP},
especially that (\ref{eq:qmnW}) and (\ref{eq:qnW})  are not direct generalizations of (\ref{eq:qmn1}) and (\ref{eq:qn1}).

Similarly to the classical case, a $\delta$-rule based BP$_4$ decoding algorithm with a serial schedule,
referred to as \emph{serial BP$_4$}, is given in Algorithm~\ref{alg:SQBP}.
Again, serial BP$_4$ and parallel BP$_4$ have the same computational complexity in an iteration.

The update schedule of BP could affect the convergence behavior a lot for quantum stabilizer codes. 
We provide two examples as follows.
First, the well-known $[[5,1]]$ code \cite{LMPZ96} can correct an arbitrary weight-one error and has a check matrix $$ S = \bsmtx{X&Z&Z&X&I\\ I&X&Z&Z&X\\ X&I&X&Z&Z\\ Z&X&I&X&Z}.$$
The Tanner graph of this matrix obviously has many four-cycles.
Parallel BP$_4$ can decode all the weight-one errors, except for the error $IIIYI$, and 
the output will oscillate continuously as shown in Fig.~\ref{fig:Y4}\,(a). However,  serial BP$_4$ converges very soon, as shown in Fig.~\ref{fig:Y4}\,(b).

	\begin{figure*}[htbp] \centering \includegraphics[width=1.0\textwidth]{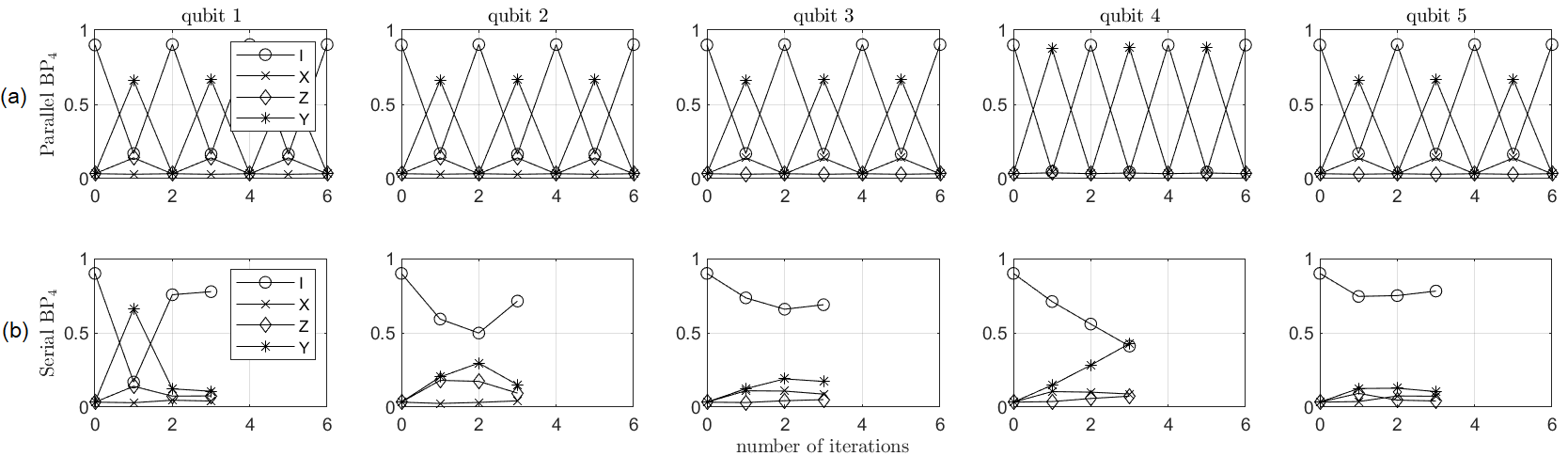}	
	\caption{The change of the probabilities of $I,X,Y,Z$, when decoding the $[[5,1,3]]$ code with error $IIIYI$ at a depolarizing error rate $\eps=0.1$}	
	\label{fig:Y4} 
	\end{figure*}

Second, 
we construct a $[[129,28]]$ hypergraph-product code \cite{TZ14} by two BCH codes with parameters $[7,4,3]$ and $[15,7,5]$, as in \cite{LP19}. 
This hypergraph-product code  {has   minimum distance $d_\text{min}=3$} and also corrects an arbitrary weight-one error.
Serial BP$_4$ greatly improves the performance of parallel BP$_4$  as shown in Fig.~\ref{fig:hyper}. 
Each data point is obtained by collecting at least 100 logical errors.
  By  mapping the $M\times N$ check matrix $S$ to an $M\times 2N$ binary matrix $H$~\cite{GotPhD,NC00},
  one can also decode the quantum code using the algorithms in Sec.~\ref{sec:ClsBP} but with  the probabilities $(p_n^{(0)},p_n^{(1)})$  initialized to $p_n^{(1)} = 2\eps/3$ (cf. (40) in \cite{MMM04}).
The performance of using BP$_2$ is also provided in Fig.~\ref{fig:hyper}. 
For reference,  the estimated performance curves of bounded distance decoding  {(BDD)} for minimum distance~3 or~5 are also provided.
BDD has an error-correction radius $t=\lfloor \frac{d_\text{min}-1}{2} \rfloor$. 
Gallager estimated that BP can correct most errors in a radius of $2t$~\cite{Gal63} (which corresponds to $d_\text{min}=5$ here). 
Using the serial schedule can achieve a performance near Gallager's estimation.

	\begin{figure}[htbp] \centering \includegraphics[width=0.5\textwidth]{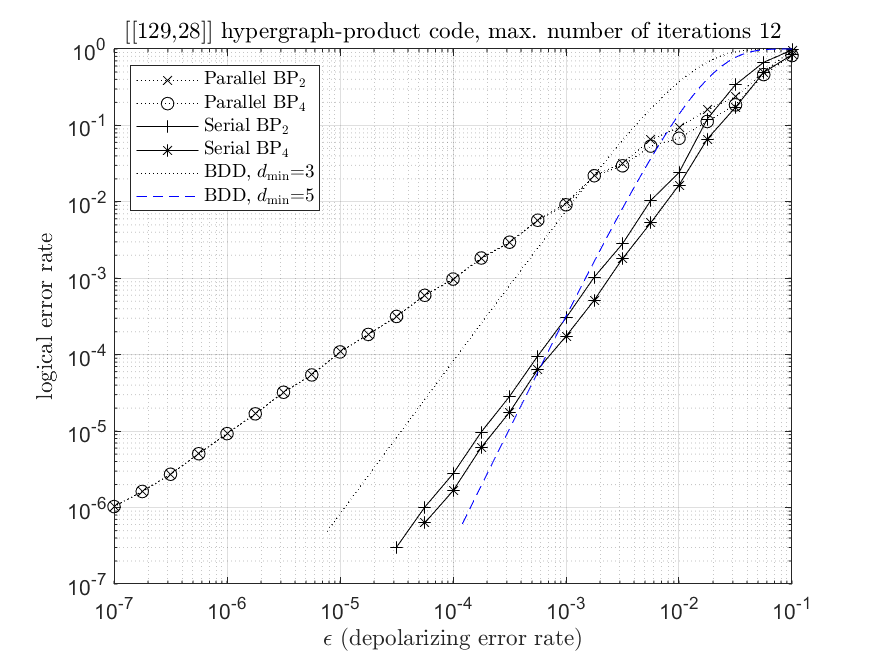}
	\caption{Performance of decoding the $[[129,28]]$ hypergraph-product code} \label{fig:hyper}
	\end{figure}


\section{Message Normalization and Offset} \label{sec:NorQBP}

In classical BP decoding, a message often has over-estimated strength  due to approximation 
(by the min-sum algorithm) \cite{CF02b,CDE+05} or cycles in the graph \cite{YHB04}. 
Normalizing or offsetting the message magnitude can be helpful.
Since our $\delta$-rule based BP algorithms for quantum stabilizer codes have only single-valued messages,
the techniques of message normalization and offset can be effortlessly applied to our algorithms. 
By viewing BP as a gradient descent algorithm \cite{LBB98}, 
one can think of message normalization/offset as adjusting the update step-size \cite{JN06,EM06}. 
The normalization/offset is defined in logarithmic domain, but can be equivalently defined in linear domain as we will do in the following.

We review the log-domain computation first. Assume that the log-likelihood ratio (LLR) $$\La = \ln\frac{P(E=0)}{P(E=1)}$$ is to be passed.
\begin{itemize}
\item {\bf Message normalization}: The message is normalized as $Q = \La/\af$ by some positive~$\af$ before passing.
\item {\bf Message offset}: The message magnitude is offset by some positive~$\beta$:
	\beq{
		Q = \bcase{	0,							 	&\text{if $|\La|<\beta$},	\\
					\sgn(\La)(|\La|-\beta),			&\text{otherwise}.			}
	}
\end{itemize}
Since $\beta$ serves as a soft threshold, the performance of message offset is usually worse \cite{CDE+05,YHB04},
but it has a lower complexity without the multiplication in message normalization. 

Now we propose various BP decoding algorithms using message normalization/offset. 
First, a BP$_4$ decoding with check-node messages normalized by parameter~$\af_c$ is defined in Algorithm~\ref{alg:QBP-a}.
Second, a BP$_4$ decoding with variable-node messages normalized by parameter~$\af_v$ is defined in Algorithm~\ref{alg:QBP-av}.
Note that the function $(\cdot)^{1/\af}$ does not need to be perfect, which  can be approximated by the method \cite{Sch99} using only one multiplication and two additions. 
Finally, a BP$_4$ decoding with check-node messages offset by parameter~$\beta$ is defined in Algorithm~\ref{alg:QBP-b}.

\begin{algorithm}[htbp] \caption{: BP$_4$ decoding with check-node messages normalized by parameter $\af_c$} \label{alg:QBP-a}
The algorithm is identical to Algorithm~\ref{alg:QBP} (parallel BP$_4$) or Algorithm~\ref{alg:SQBP} (serial BP$_4$) except that
$r_{mn}^{(0)}$ and $r_{mn}^{(1)}$ are replaced by $r_{mn}^{(0)} = (\frac{1+\dlt_{mn}}{2})^{1/\af_c}$ and $r_{mn}^{(1)} = (\frac{1-\dlt_{mn}}{2})^{1/\af_c}$, respectively, for some $\af_c>0$. 
\end{algorithm}

\begin{algorithm}[htbp] \caption{: BP$_4$ decoding with variable-node messages normalized by parameter $\af_v$} \label{alg:QBP-av}
The algorithm is identical to Algorithm~\ref{alg:QBP} (parallel BP$_4$) or Algorithm~\ref{alg:SQBP} (serial BP$_4$) except that  
$q_{mn}^{(0)}$ and $q_{mn}^{(1)}$ are replaced by 
$q_{mn}^{(0)}=a_{mn}(q_{mn}^I+q_{mn}^{S_{mn}})^{1/\af_v}$ and $q_{mn}^{(1)}=a_{mn}(\sum_{W'\in\{X,Y,Z\}\setminus S_{mn}}q_{mn}^{W'})^{1/\af_v}$, respectively, 
for some $\af_v>0$, where   $a_{mn}$ is a chosen scalar such that $q_{mn}^{(0)}+q_{mn}^{(1)}=1$.
\end{algorithm}

\begin{algorithm}[htbp] \caption{: BP$_4$ decoding with check-node messages offset by parameter $\beta$} \label{alg:QBP-b}
The algorithm is identical to Algorithm~\ref{alg:QBP} (parallel BP$_4$) or Algorithm~\ref{alg:SQBP} (serial BP$_4$) except that 
$r_{mn}^{(0)}$ and $r_{mn}^{(1)}$ are computed accordingly as follows: 
  \beqs{
	&\text{ If $r_{mn}^{(0)}/r_{mn}^{(1)} > e^\beta$, update $r_{mn}^{(0)}$ to $r_{mn}^{(0)} / e^\beta$; } \\
	&\text{ if $r_{mn}^{(1)}/r_{mn}^{(0)} > e^\beta$, update $r_{mn}^{(1)}$ to $r_{mn}^{(1)} / e^\beta$; } \\
	&\text{ otherwise, set both $r_{mn}^{(0)}$ and $r_{mn}^{(1)}$ to $1/2$. }
  }
\end{algorithm}

To sum up, we have the following algorithms:
\begin{itemize}
	\item Parallel BP$_4$: Algorithm~\ref{alg:QBP}.
	\item Serial BP$_4$: Algorithm~\ref{alg:SQBP}.
	\item Parallel/Serial BP$_4$, normalized by $\af_c$: Algorithm~\ref{alg:QBP-a}.
	\item Parallel/Serial BP$_4$, normalized by $\af_v$: Algorithm~\ref{alg:QBP-av}.
	\item Parallel/Serial BP$_4$, offset by $\beta$: Algorithm~\ref{alg:QBP-b}.
\end{itemize}
We evaluate the error performance  and complexity (in terms of average number of iterations) of these algorithms by simulating quantum bicycle codes \cite{MMM04}. 
The check matrix $H$ of a quantum bicycle code is constructed by concatenating a random sparse circulant matrix $C$ and its transpose and then deleting some rows. 
%
%
For every data point in the simulation, at least 100 logical errors are collected.
The degeneracy is considered in decoding: 
Decoding is successful when the estimated error is equivalent to the actual error  up to  some stabilizer. 
Otherwise, a logical error occurs, which could be either a detected error
or an undetected error.

First, we consider a $[[256,32]]$ quantum bicycle code  {with a check matrix generated by a random circulant matrix of row-weight $8$} (the same parameters as an example in \cite{LP19}).
Although the beliefs may propagate faster in serial BP$_4$, the effects of wrong beliefs may cause worse performance.
Since wrong beliefs mostly come from   over-estimated messages due to short cycles, suppressing the message strength could be helpful. 
We apply the normalization/offset methods (by $\af_c$,~$\af_v$, and~$\beta$, respectively) and the performance is improved significantly, 
as shown in Figures~\ref{fig:256_ac},~\ref{fig:256_av}, and~\ref{fig:256_b}, respectively. 
Stronger message (smaller~$\eps$) would need larger suppression (larger $\af_c$,~$\af_v$, or~$\beta$). 
It is possible to choose different~$\af_c$,~$\af_v$, or~$\beta$ for different~$\eps$ to achieve a better performance.
The decoding complexity (evaluated by the average number of iterations) is shown in
Figures~\ref{fig:256it_ac},~\ref{fig:256it_av}, and~\ref{fig:256it_b}, respectively. 
The $\af_c$ method has a lower complexity. The $\af_v$ method achieves a (slightly) better performance at the cost of higher complexity.
The $\beta$ method needs a careful selection of the value of $\beta$ (due to the threshold effect), 
though the final performance is not as good as the normalization method (as expected). 
Its possible advantage is the save of multiplication if  our procedure can be transformed to work in log-domain \cite{WSM04}.

Next, we consider an $[[800,400]]$ quantum bicycle code  {with a check matrix generated by a random circulant matrix of row-weight $15$} (the same parameters as an example in \cite{PC08}).
The two normalization methods (by $\af_c$ and $\af_v$ respectively) again greatly improve the performance  (especially the $\af_v$ method), 
as shown in Figures~\ref{fig:800_ac} and~\ref{fig:800_av}. 
Serial BP$_4$ performs much better now, though it still hits a high error floor. 
 Both normalization methods improve the error floor performance a lot.
 Even the worse parallel BP$_4$ has improved error floor performance.
 This is very useful in practice for full parallelism.
The decoding complexity is shown in Figures~\ref{fig:800it_ac} and~\ref{fig:800it_av}, respectively, in terms of average number of iterations.
The~$\af_v$ methods improve the performance further at the cost of higher complexity.
Applying $\af_c$ (or $\af_v$) sometimes has a lower average number of iterations compared to no message normalization.
However, it only means a faster convergence rather than a lower complexity, since the normalization requires additional multiplications.


We can also use BP$_2$ to decode the quantum stabilizer codes as in the previous section. 
The message normalization can be similarly applied in BP$_2$. 
We compare BP$_2$ and BP$_4$ with the $\alpha_v$ method on the $[[256,32]]$ code  and the results are shown in Fig.~\ref{fig:256_GF2_av}.
An expectation in \cite{DM98} is that, for the same channel, 
a higher-order GF($q$) code should perform better (assuming that there are not many short cycles). 
We found a  result of contrary that, before applying any message normalization,    BP$_4$   performs worse than  BP$_2$  due to the many short cycles (though BP$_4$ performs better than BP$_2$ for larger $\eps$). 
After applying the message normalization (to suppress wrong beliefs),  BP$_4$ can outperform BP$_2$, much matching the expectation in \cite{DM98}. 
%

	\begin{figure}[htbp]
	\centering \includegraphics[width=0.5\textwidth]{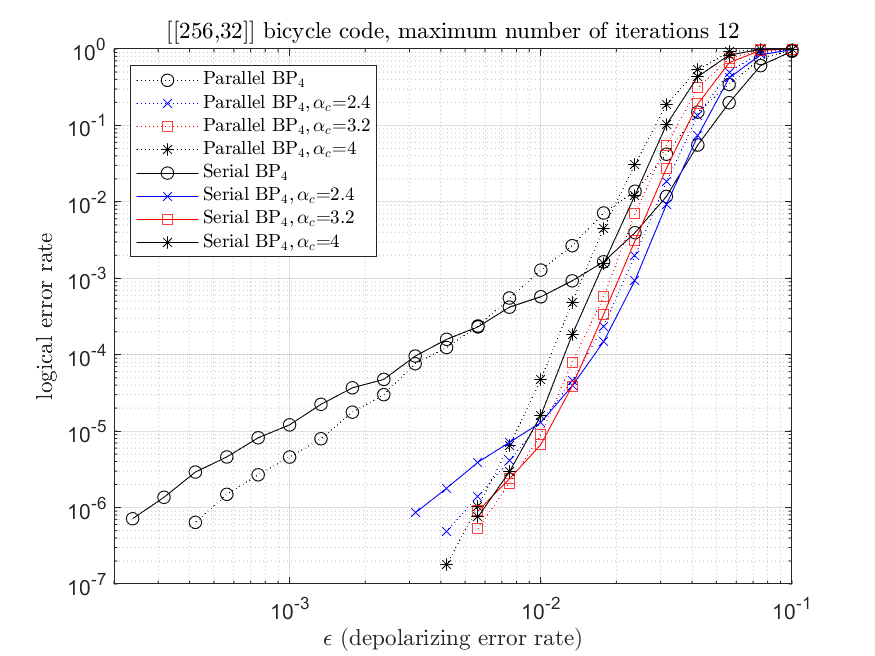}
	\caption{Performance of decoding the $[[256,32]]$ code by different $\af_c$} \label{fig:256_ac}		\vspace*{\floatsep}
	\centering \includegraphics[width=0.5\textwidth]{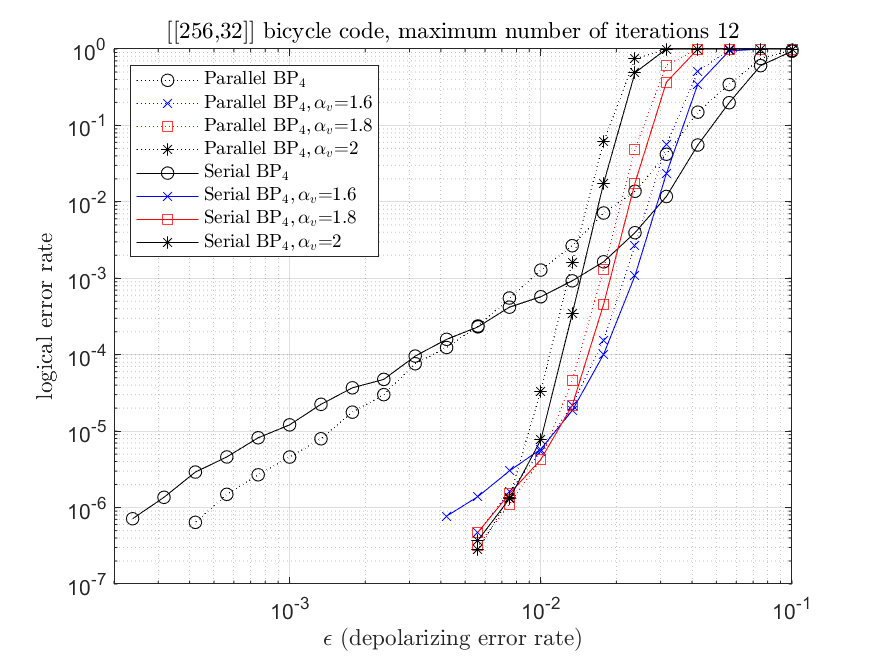}
	\caption{Performance of decoding the $[[256,32]]$ code by different $\af_v$} \label{fig:256_av}		\vspace*{\floatsep}
	\centering \includegraphics[width=0.5\textwidth]{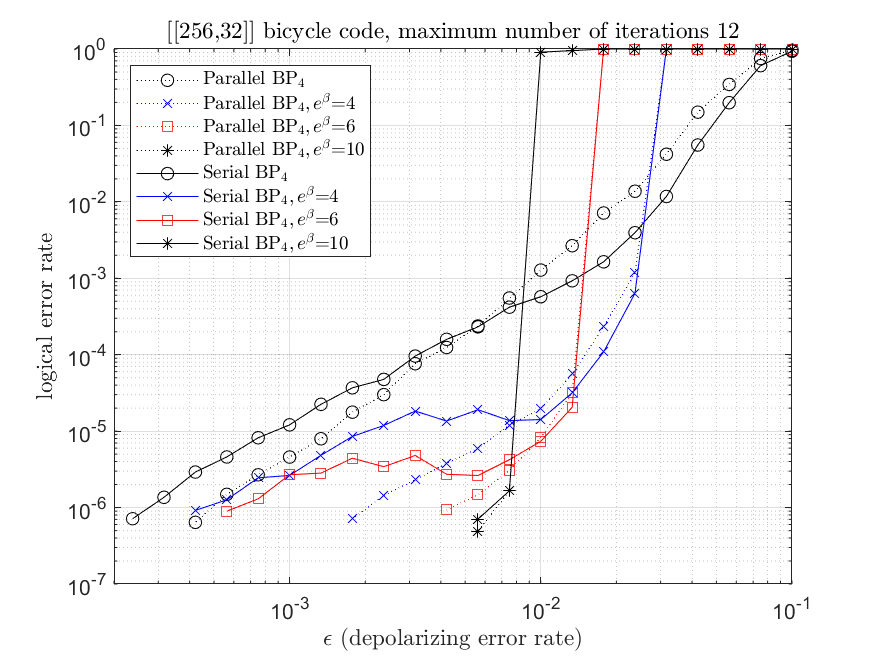}
	\caption{Performance of decoding the $[[256,32]]$ code by different $\beta$} \label{fig:256_b}
\end{figure}

\begin{figure}[htbp]
	\centering \includegraphics[width=0.5\textwidth]{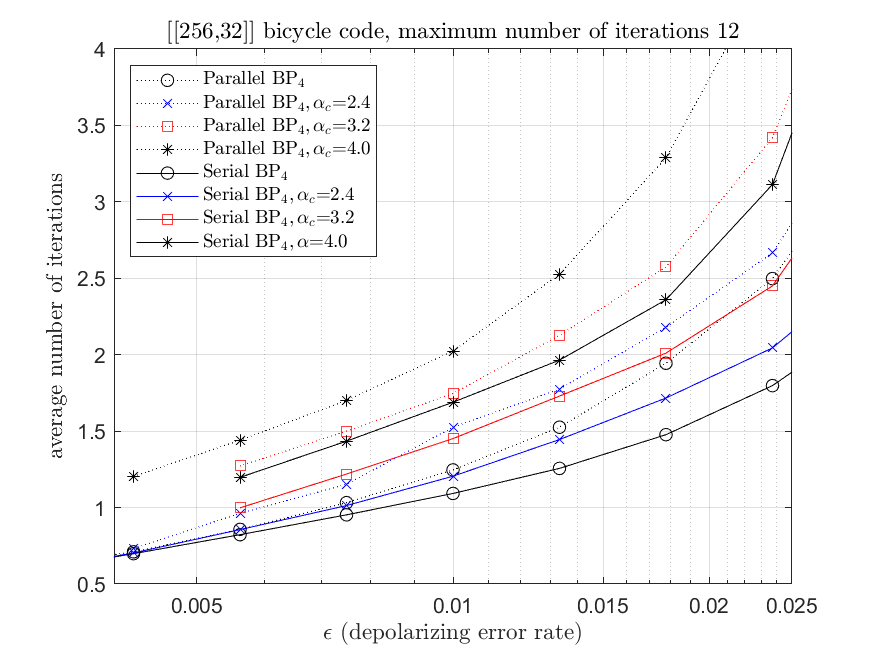}
	\caption{Complexity of decoding the $[[256,32]]$ code by different $\af_c$} \label{fig:256it_ac}	\vspace*{\floatsep}
	\centering \includegraphics[width=0.5\textwidth]{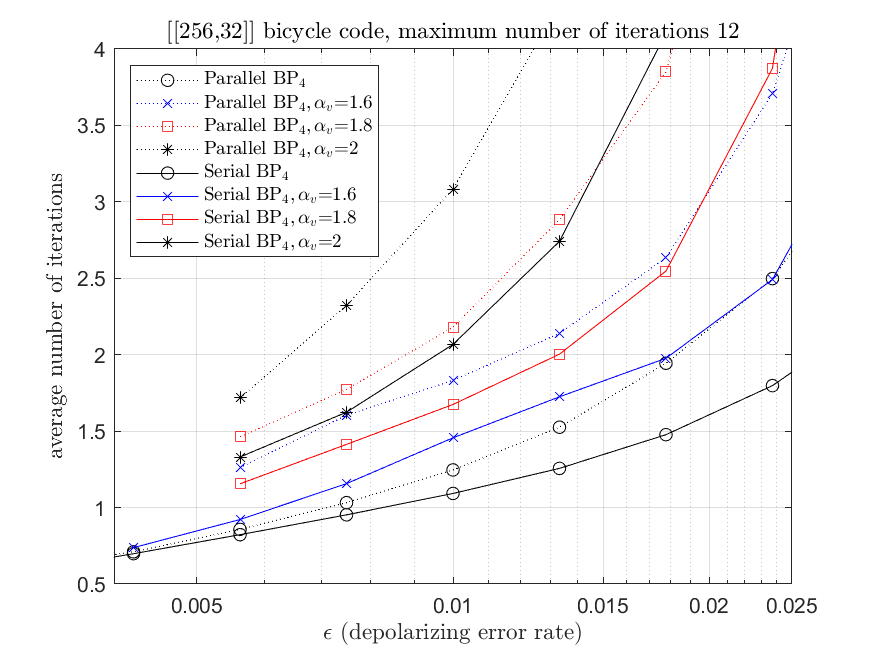}
	\caption{Complexity of decoding the $[[256,32]]$ code by different $\af_v$} \label{fig:256it_av}	\vspace*{\floatsep}
	\centering \includegraphics[width=0.5\textwidth]{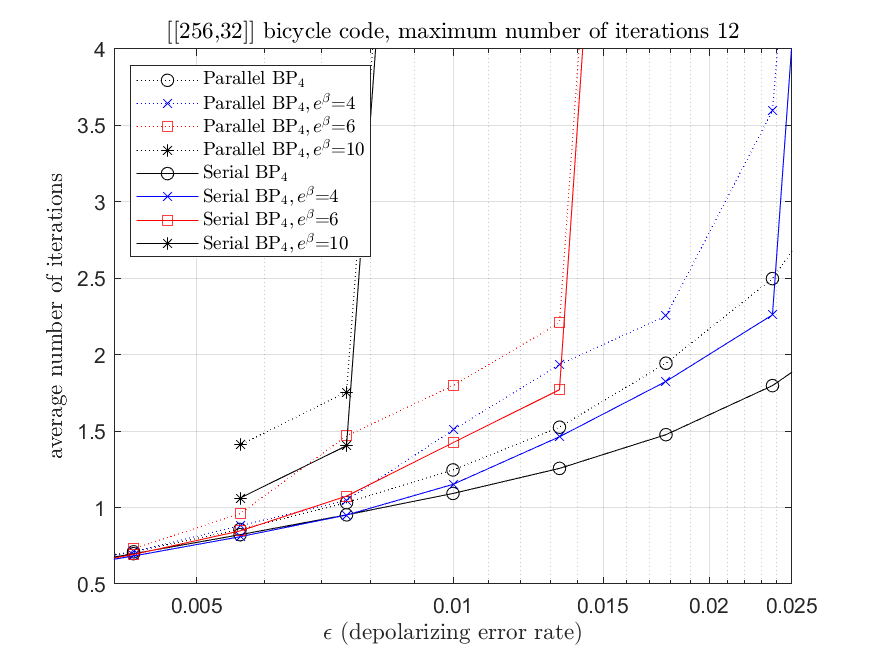}
	\caption{Complexity of decoding the $[[256,32]]$ code by different $\beta$} \label{fig:256it_b}
\end{figure}

	\begin{figure}[htbp] 
	\centering \includegraphics[width=0.5\textwidth]{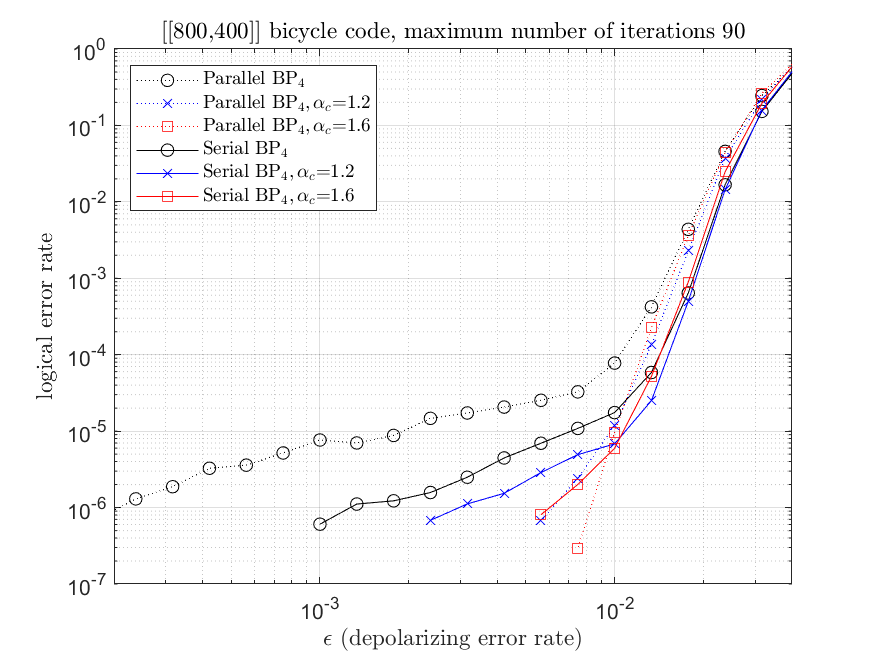}
	\caption{Performance of decoding the $[[800,400]]$ code by different $\af_c$} \label{fig:800_ac}	\vspace*{\floatsep}
	\centering \includegraphics[width=0.5\textwidth]{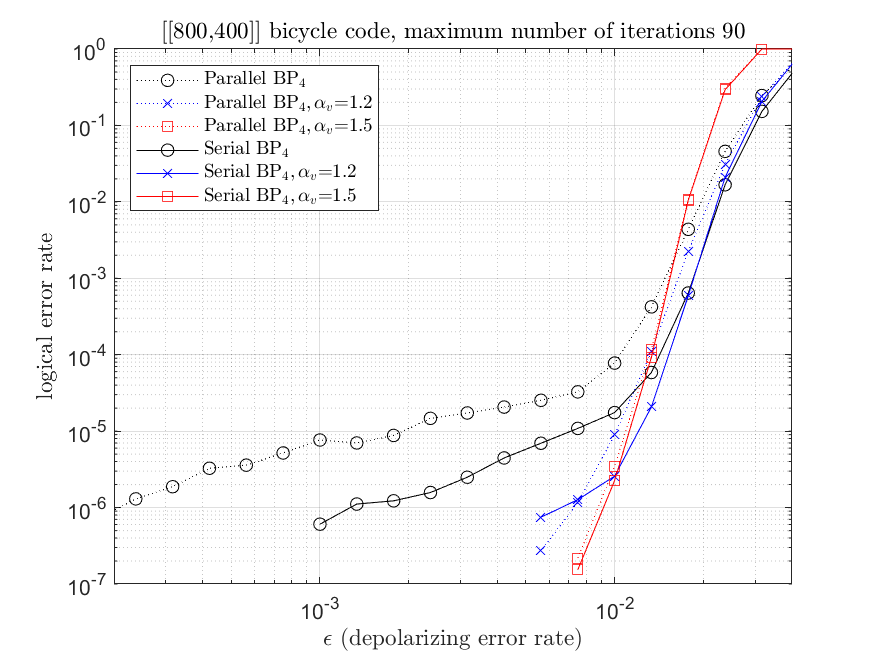}
	\caption{Performance of decoding the $[[800,400]]$ code by different $\af_v$} \label{fig:800_av}
	\end{figure}

	\begin{figure}[htbp]
	\centering \includegraphics[width=0.5\textwidth]{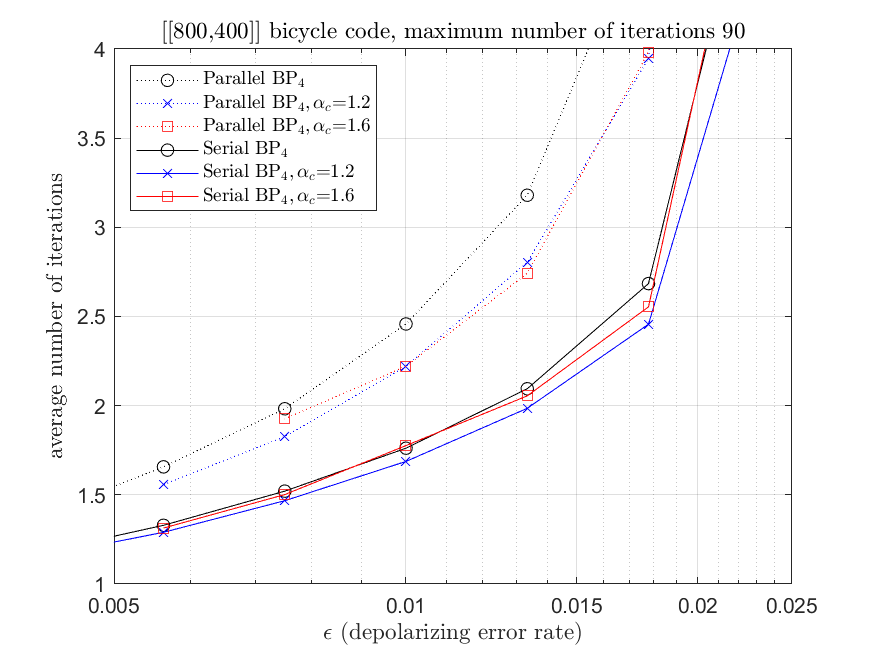}
	\caption{Complexity of decoding the $[[800,400]]$ code by different $\af_c$} \label{fig:800it_ac}	\vspace*{\floatsep}
	\centering \includegraphics[width=0.5\textwidth]{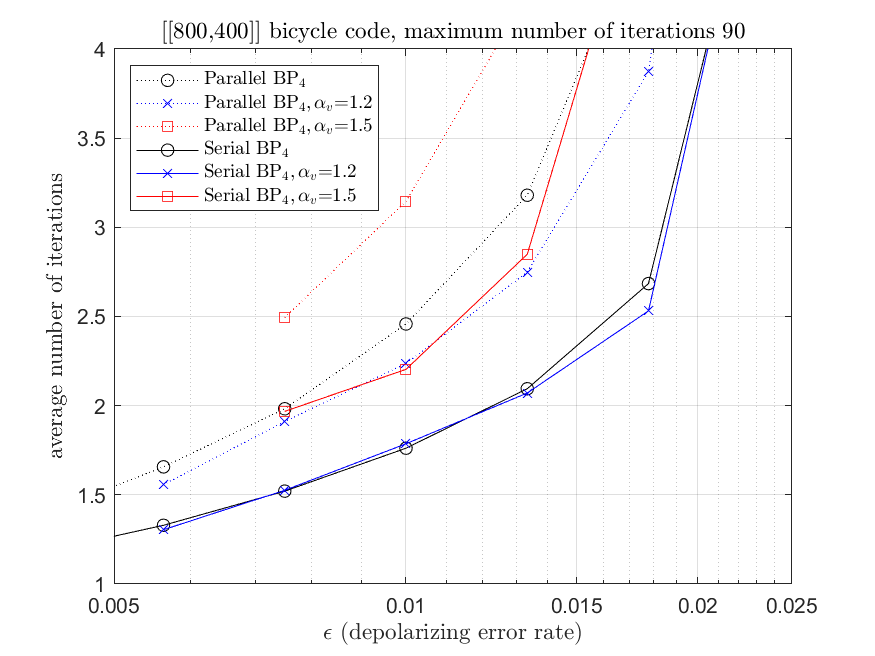}
	\caption{Complexity of decoding the $[[800,400]]$ code by different $\af_v$} \label{fig:800it_av}	\vspace*{\floatsep}	\vspace*{\floatsep}	
	\end{figure}

	\begin{figure}[htbp]
	\centering \includegraphics[width=0.5\textwidth]{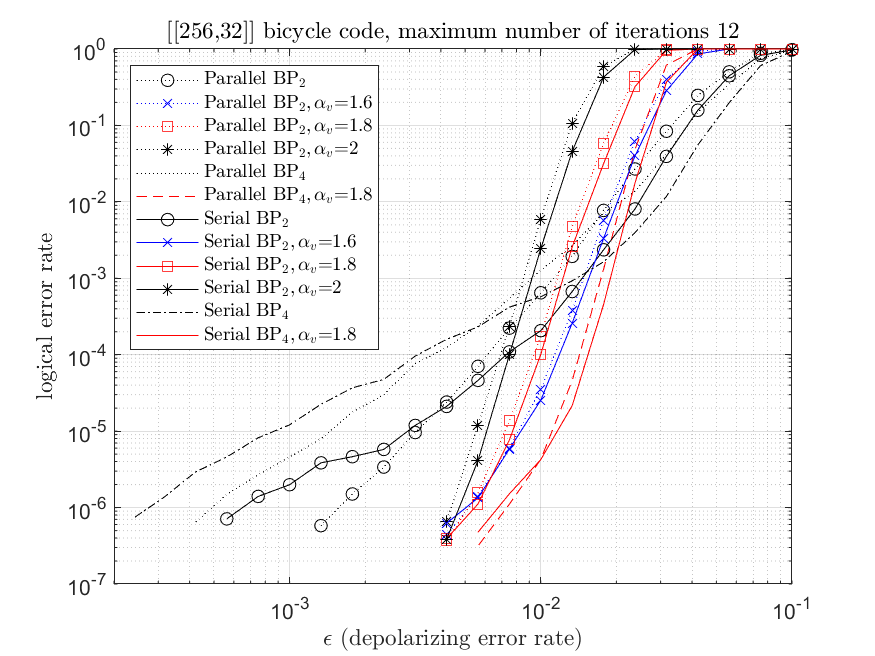}
	\caption{Performance of decoding the $[[256,32]]$ code by different $\af_v$, with BP$_2$/BP$_4$ compared.} \label{fig:256_GF2_av}
	\end{figure}

\section{Conclusion and Future Work} \label{sec:Conclu}

We proposed a refined BP decoding algorithm for quantum stabilizer codes. 
Using $\dlt$-rule to pass single-valued messages improves the check-node efficiency by a factor of~16. 
The single-valued messages can be normalized to improve the decoding performance, which works for different update schedules.
To have further improvement, additional processes (such as  {heuristic/feedback/training/redundant checks/OSD \cite{PC08,Wan+12,Bab+15,LP19,ROJ19,PK19}}) 
could be incorporated if the complexity is affordable. 
For any improvement, the efficiency is concerned since the coherence of quantum states decays very fast.

We considered parallel and serial schedules.
It may be worth to apply/develop other fixed or dynamic (adaptive) schedules~\cite{CGW07}. 

It is straightforward to transform our single-valued message-passing algorithm to a min-sum algorithm (further lower complexity) \cite{Wib96},  
but the message approximation and compensation would be more challenging due to short cycles \cite{CF02b,CDE+05,YHB04}.
It might be possible to transform our procedure to log-domain \cite{WSM04} to lower the message-offset complexity. 
It is also possible to generalize our approach to design a BP algorithm with single-valued messages for quantum codes over GF($q=2^m$) for  even $m$ since the syndromes are binary.
 We have been working on this generalization.

For the simulations in this paper, all the observed logical errors are detected errors. 
On the other hand, when a decoding is successful, we only observed that the output error is exactly the actual error.
So there may be some room for improving the decoder by exploiting the code degeneracy. 
	However, if a decoder performs well, the room for improvement could be small~\cite{BSV14}. 
The BP performance diverges badly for bicycle codes with small row-weight \cite{MMM04}. 
For this case, exploiting the degeneracy may help and we have some ongoing work.

An interesting question is to decode topological codes using our methods.
It is plausible to apply our procedure to more sparse quantum codes \cite{COT05,HI07,Djo08,Aly08,KHIS11,TZ14,TL09,CDZ13,KP13}.
Since sparse topological codes {(such as the toric code \cite{Kit03})} may have high degeneracy, we need to consider how to take advantage of the code degeneracy in BP and this is our ongoing work.



%
%
%

\section*{Acknowledgment}
CYL was financially supported from the Young Scholar Fellowship Program by Ministry of Science and Technology (MOST) in Taiwan, under Grant MOST108-2636-E-009-004.
The authors would like to thank four anonymous reviewers for their valuable and constructive comments on our manuscript.

\ifCLASSOPTIONcaptionsoff
  \newpage
\fi

\bibliographystyle{IEEEtran}


\end{document}